\RequirePackage{luatex85}
\documentclass[]{aa} 
\usepackage{color, soul}
\usepackage{graphicx}
\usepackage{txfonts}
\usepackage{subfigure}
\usepackage{cases}
\usepackage{gensymb}
\usepackage{changepage}
\usepackage{natbib}
\usepackage{multirow}
\usepackage{hyperref}
\usepackage{caption}

\begin{document}

\title{Investigating the outskirts of Abell 133 with {\it Suzaku} and {\it Chandra} observations}

\author{Zhenlin Zhu\inst{1,2}
\and Orsolya E. Kov\'acs\inst{3}
\and Aurora Simionescu\inst{1,2,4}
\and Norbert Werner\inst{3}
}

\institute{SRON Netherlands Institute for Space Research, Niels Bohrweg 4, 2333 CA Leiden, The Netherlands\\\email{Z.Zhu@sron.nl}\label{inst1}
\and  Leiden Observatory, Leiden University, Niels Bohrweg 2, 2300 RA Leiden, The Netherlands\label{inst2} 
\and Department of Theoretical Physics and Astrophysics, Faculty of Science, Masaryk University, Kotl\'{a}rsk\'{a} 2, 61137 Brno, Czech Republic\label{inst3} 
\and  Kavli Institute for the Physics and Mathematics of the Universe, The University of Tokyo, Kashiwa, Chiba 277-8583, Japan \label{inst4}
}

\abstract{
Past observations and simulations predict an increasingly inhomogeneous gas distribution towards the outskirts of galaxy clusters, but the exact properties of such gas clumping are not yet well known. The outskirts of Abell 133 benefit from deep X-ray observations, with a 2.4 Ms ultra-deep {\it Chandra} exposure as well as eight archival {\it Suzaku} pointings, making it a unique laboratory to study the clumping of the intracluster medium.

}
{
We searched for significant clump candidates, in particular aiming to identify those that could represent genuine ICM inhomogeneity.
To further understand how clumping biases the thermodynamic profiles, we compared the measurements including and excluding the clump candidates.
}
{
We jointly analyzed {\it Chandra} and {\it Suzaku} observations of Abell 133. We selected clump candidates with at least 2 $\sigma$ significance based on the {\it Chandra} image and further discussed their origins using information from the DESI Legacy Imaging Surveys cluster catalogue, as well as the CFHT r-band image. We performed multiple rounds of {\it Suzaku} spectral analysis with different corrections for the underlying point sources and clump distribution, and compared the resulting thermodynamic profiles.
}
{
We detected 16 clump candidates using {\it Chandra}, most of which are identified as background clusters or galaxies as opposed to intrinsic inhomogeneity. 
Even after the correction of the resolved clumps, the entropy profile approaching the outskirts still flattens, deviating from the power law model expected from self-similar evolution, which implies that unresolved clumping and other complex physics should contribute to the entropy flattening in the outskirts. 
}
{
}

\maketitle
   
\section{Introduction}
\label{sec:intro}

The properties of the intracluster medium (ICM) in the outskirts of galaxy clusters have been actively explored with a combination of multi-wavelength observations and state-of-the-art numerical simulations (for a review, see \citealp{Walker2019} ). These outer regions of galaxy clusters are an ideal probe of the ongoing process of virialization due to large-scale structure growth. Both predictions from numerical simulations and observational studies indicate that cluster outskirts exhibit distinct marks of recent accretion from the surrounding large-scale structure, e.g. an inhomogeneous gas density distribution \citep{Simionescu2011,Nagai2011}, non-thermal pressure/turbulence \citep{Ghirardini2018,Nelson2014} and non-equilibrium electrons in the ICM \citep{Hoshino2010,Avestruz2015}. 
An increase of the gas clumping at larger radii can lead to significant systematic biases in the X-ray measurements of various ICM properties, for instance to overestimations of the density and, consequently, underestimations of the gas entropy. Although this effect has been invoked in numerous works to explain why the observed entropy profiles deviate from the expected baseline derived from pure gravitational heating \citep[e.g.][]{Walker2013,Urban2014,Tchernin2016,Simionescu2017}, as mentioned above, several other physical effects can simultaneously be at play in addition to the gas clumping. Only deep observations offering both sharp imaging and reliable spectral information of these faint regions can help us to truly disentangle the complex physics of cluster outskirts.



To date, {\it Chandra} has accumulated ultra-deep exposures on Abell 133 and Abell 1795, out to very large radii, providing a special coverage in the outskirts even beyond $r_{200}$.
{\it Suzaku} observations offer a similar coverage out to $r_{200}$, enabling us to directly measure the thermodynamic properties owing to the relatively low and stable background.
This unique combination of available data makes these clusters ideal probes for understanding the complex physics of the ICM near the virial radii.

Abell 133 is an X-ray luminous cool-core galaxy cluster at z=0.0566 \citep{Struble1999}, which has a cool core and a central radio source.
Previous {\it Chandra} studies with relatively shallow exposure (\citealp{Fujita2002}; \citealp{Randall2010}) showed that the cluster core of Abell 133 has a complex morphology, and prominent radio relics indicative of ongoing merger activity.
Using the 2.4 Ms deep {\it Chandra} observations, \citet{Vikhlinin2013} reported the detection of three X-ray filaments extending outside of $r_{200}$.
To verify this claim, \citet{Connor2018} conducted a spectrographic campaign on the Baade 6.5m telescope around Abell 133 and found corresponding galaxy enhancements along those filaments.
With this ultra-deep dataset, \citet{Morandi2014} have shown that clumping increases with radius, however, they did not study the nature of individual clumps.

In this paper, we report on results from a combined X-ray analysis using {\it Chandra} and {\it Suzaku} observations of Abell 133.
To precisely correct the thermodynamic profiles for this gas clumping effect, we attempt to identify significant clumps based on {\it Chandra} images and mask them in the {\it Suzaku} spectral analysis.
We introduce the imaging analysis in Section \ref{sec:imaging} and present surface brightness profiles in Section \ref{sec:sb}.
We list the 16 {\it Chandra}-selected clump candidates in Section \ref{sec:clump}, and discuss their origins.
Our {\it Suzaku} spectral analysis methods are described in Section \ref{sec:spec}. 
Section \ref{sec:thermo} shows the resulting thermodynamic profiles.
Furthermore, in Section \ref{sec:discussion}, we discuss how the profiles are affected by clumping corrections, as well as other systematic uncertainties which can be mitigated owing to the combination of {\it Chandra} and {\it Suzaku} coverage.
In a companion paper (Kovács et al. 2023 submitted), we applied the same techniques for Abell 1795 and discussed the corresponding results.
In this work, we adopt a $\Lambda$CDM cosmology with $\Omega_{m}$=0.27 and H$_{0}$ = 70 ${{\rm km}~{\rm s}^{-1}{\rm Mpc}^{-1}}$.
This gives a physical scale 1{\arcmin} =  64.2 kpc for A133 cluster at redshift {\it z} = 0.0566.
Our reference values for r$_{500}$ and r$_{200}$ are 14.6{\arcmin} and 24.7{\arcmin} respectively \citep{Randall2010}. 
All errors are given at the 68\% confidence level unless otherwise stated.

\section{Observations and data reduction}
\label{sec:data}
\subsection{{\it Suzaku} }
We utilized eight {\it Suzaku} observations mapping the Abell 133 cluster to 1.2$r_{200}$.
The four observations (E, S, N, W) pointed to the inner core region have been studied in \citep{Ota2016}, while the later observations towards the outskirts have not been explored so far.
Detailed information can be found in Table \ref{tab:suzaku}.

The X-ray Imaging Spectrometer(XIS) data were analyzed following the procedure described in \citet{Simionescu2013}, \citet{Urban2014} and \citet{Zhu2021}.
In brief, we used the cleaned events files produced by the standard screening process\footnote{https://heasarc.gsfc.nasa.gov/docs/suzaku/analysis/abc/node9.html}, and applied the following additional filtering criteria.
The observation periods with low geomagnetic cut-off rigidity (COR $\leq$ 6 GV) were excluded.
For observations later than 2011, we excluded two columns on either side of the charge-injected columns for the XIS1 detector, to avoid the charge leak effect.
The vignetting effect has been corrected using ray-tracing simulations of extended, spatially uniform emission.
The data reduction was performed with HEAsoft v6.26.
We have examined the 0.7-3 keV light curves of each observation with a time bin of 256 seconds, to further ensure no flaring occurred during the clean exposure.
In addition, we checked for potential contamination from solar wind charge exchange (SWCX), by plotting the proton flux measured by the {\it Wind} spacecraft's solar wind experiment instrument\footnote{https://wind.nasa.gov/}, as shown in Figure \ref{fig:swce}. 
We found the proton flux curves of the eight observations are much lower than 4$\times$ 10$^{8}$ cm$^{-2}$ s$^{-1}$, therefore the contamination from geocoronal SWCX is negligible \citep{Yoshino2009}.

\begin{table}
\centering
\caption{{\it Suzaku} Observational Log}
\begin{tabular}{l c c c}
\hline\hline
Observation & OBSID & Start Date & Exposure (ks) \\[5pt]
E & 805021010 & 2010 Jun 9 & 51.6 \\
S & 805022010  & 2010 Jun 8  & 51.1 \\
N & 805020010 & 2010 Jun 5  & 50.2 \\
W & 805019010  & 2010 Jun 7 & 50.0 \\
Field 1& 808081010 & 2013 Dec 19 & 53.7 \\
Field 2 & 808082010 & 2013 Dec 20  & 50.7 \\
Field 3 & 808083010 & 2013 Dec 5 & 51.9 \\
Field 4 & 808084010 & 2013 Dec 6 & 52.5 \\\hline
\end{tabular}
\label{tab:suzaku}
\vspace{10pt}
\end{table}

\begin{table}
\begin{center}
\caption{{\it Chandra} Observational Log}
\label{tab:chandra_log}
\begin{tabular}{l c c c c}
\hline\hline
Obs ID  & Start Date & Exposure & R.A. & Dec. \\[5pt]
 & & (ks) & (deg) & (deg) \\\hline
  &   &  & & \\
3183   & 2002  Jun  24   &   38.01 &  15.459 & -21.817 \\ 
3710   & 2002  Jun  26   &   37.80 &  15.459 & -21.817 \\ 
9897   & 2008  Aug  29   &   61.89 &  15.680 & -21.865 \\ 
12177  & 2010  Aug  31   &   46.95 &  15.459 & -22.063 \\ 
12179  & 2010  Sep  03   &   44.51 &  15.741 & -22.143 \\ 
12178  & 2010  Sep  07   &   42.46 &  16.005 & -22.005 \\ 
13391  & 2011  Aug  16   &   39.19 &  15.759 & -22.229 \\ 
13442  & 2011  Aug  23   &  152.74 &  15.353 & -21.813 \\ 
13443  & 2011  Aug  26   &   60.30 &  15.361 & -21.808 \\ 
14333  & 2011  Aug  31   &  114.83 &  15.776 & -21.613 \\ 
13445  & 2011  Sep  02   &   55.60 &  15.515 & -22.191 \\ 
13444  & 2011  Sep  03   &   32.41 &  15.776 & -21.611 \\ 
13449  & 2011  Sep  06   &   56.74 &  15.426 & -21.661 \\ 
12178  & 2010  Sep  07   &   42.46 &  16.005 & -22.005 \\ 
13447  & 2011  Sep  08   &   57.62 &  15.985 & -21.744 \\ 
13446  & 2011  Sep  09   &   49.59 &  15.482 & -21.585 \\ 
14338  & 2011  Sep  10   &  100.87 &  15.482 & -21.585 \\ 
14343  & 2011  Sep  12   &   30.48 &  15.995 & -21.708 \\ 
13448  & 2011  Sep  13   &  126.63 &  15.995 & -21.709 \\ 
13392  & 2011  Sep  16   &   42.80 &  15.642 & -21.562 \\ 
13518  & 2011  Sep  17   &   43.78 &  15.742 & -21.899 \\ 
13454  & 2011  Sep  19   &   79.55 &  16.033 & -21.896 \\ 
14346  & 2011  Sep  21   &   72.02 &  16.033 & -21.897 \\ 
14345  & 2011  Sep  23   &   30.62 &  15.324 & -22.089 \\ 
13452  & 2011  Sep  24   &  127.16 &  15.324 & -22.088 \\ 
13450  & 2011  Oct  05   &   94.85 &  15.980 & -22.160 \\ 
14347  & 2011  Oct  09   &   59.23 &  15.977 & -22.162 \\ 
14354  & 2011  Oct  10   &   35.43 &  15.638 & -22.149 \\ 
13453  & 2011  Oct  13   &   60.48 &  15.843 & -21.606 \\ 
13456  & 2011  Oct  15   &  124.10 &  15.638 & -22.149 \\ 
13455  & 2011  Oct  19   &   60.55 &  15.984 & -21.917 \\ 
13457  & 2011  Oct  21   &   59.83 &  15.341 & -22.088 \\ 
20786*  & 2018  Oct  01   &   20.13 &  16.059 & -22.404 \\ 
21864*  & 2018  Oct  05   &   31.03 &  16.059 & -22.404 \\ 
21865*  & 2018  Oct  06   &   35.00 &  16.058 & -22.404 \\ 
20787*  & 2018  Oct  07   &   35.50 &  16.193 & -22.289 \\ 
21872*  & 2019  Feb  07   &   25.01 &  16.279 & -22.596 \\ 
21873*  & 2019  Feb  08   &   28.28 &  16.295 & -22.357 \\ 

\hline
\end{tabular}
\end{center}
\tablefoot{The R.A. and Dec. columns indicate the pointing centers of each Chandra observation. The observations denoted with asterisks are not included in our main analysis. }
\end{table}

\subsection{\it Chandra}

Abell 133 was frequently observed by {\it Chandra}'s Advanced CCD Imaging Spectrometer (ACIS) from 2002 to 2019.
We utilized 38 observations taken in ACIS-I VFAINT mode, which accumulated to a total of 2.26 Ms clean exposure (See Table \ref{tab:chandra_log}).

All data were reprocessed using the CIAO data analysis package, version 4.13, and the latest calibration database (CALDB 4.9.4) distributed by the Chandra X-ray Observatory Center.
We removed the cosmic rays and bad pixels from all level-1 event files using the CIAO tool {\it chandra\_repro} with VFAINT mode background event filtering.
For each observation, we extracted the lightcurve in the 9--12 keV band, to examine possible contamination from flare events.
To remove the time intervals with anomalous background, we applied the CIAO tool {\it deflare} to filter out the times where the background rates exceed $\pm 2 \sigma$ of the average value.

We utilized stowed background files to estimate the non-X-ray background (NXB). 
All stowed background events files were combined and reprocessed with {\it acis\_process\_events} using the latest gain calibration files.
For each observation, we scaled the NXB to match the 9-12 keV count rate of the observation.

\section{X-ray Imaging}
\label{sec:imaging}
\subsection{\it Suzaku}
\begin{figure}
\centering
\includegraphics[width=0.49\textwidth]{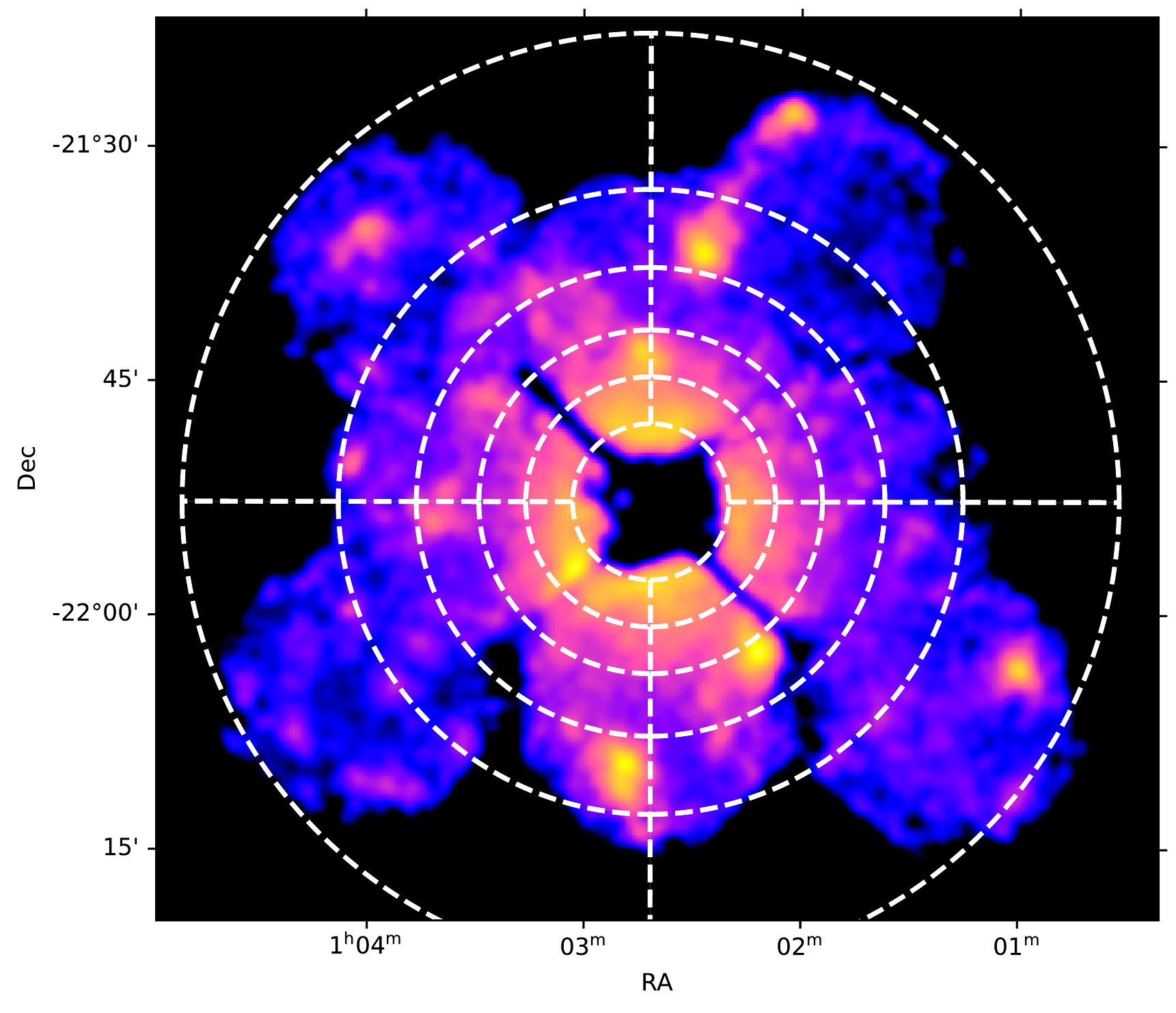}
\caption{Exposure- and vignetting-corrected 0.7-7 keV Suzaku image of Abell 133. The 90-degree white annuli show extraction regions applied in the spectral analysis. }
\label{fig:suzaku_fig}
\end{figure}

We extracted images from all three XIS detectors in the 0.7-7 keV band and removed a 30$\arcsec$ region around the detector edges. 
To minimize the influence of systematic uncertainties related to the vignetting correction, pixels with an effective area less than half of the on-axis value were also masked.
Using night Earth observations, we generated the corresponding instrumental background images. 
Vignetting effects were corrected after background subtraction using ray-tracing simulations of extended, spatially uniform emission. 
Figure \ref{fig:suzaku_fig} shows the resulting flux image of Abell 133, smoothed with a Gaussian kernel of 25$\arcsec$.
We identified nine point sources based on the {\it Suzaku} image, which are shown in Figure \ref{fig:suzaku_psrc}.

\subsection{\it Chandra}
\label{sec:chandra_imaging}
\begin{figure}
\centering
\includegraphics[width=0.49\textwidth]{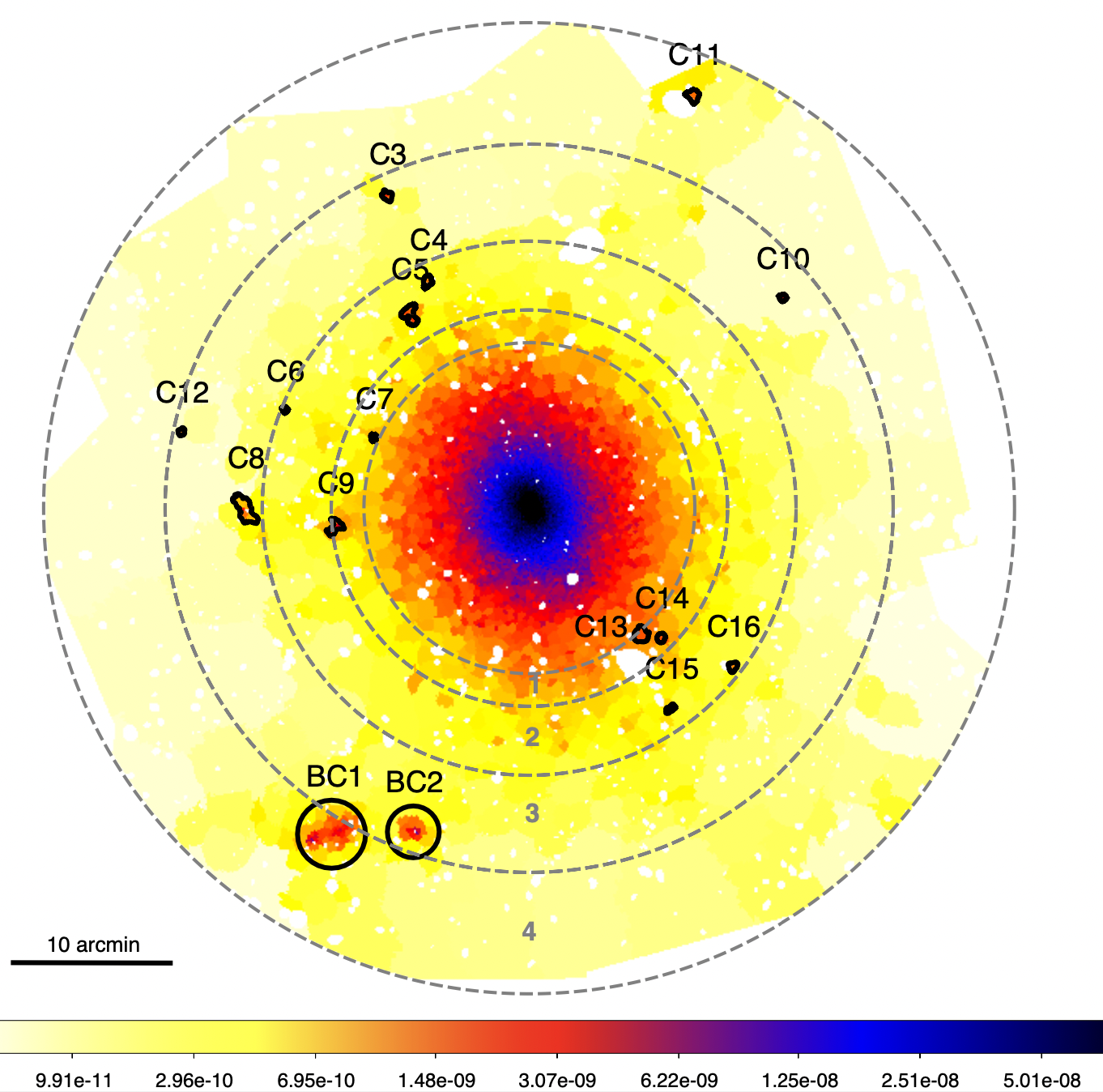}
\caption{Contour binned image of the inner 30$\arcmin$ of Abell 133 in the 0.5--3 keV energy band. The point sources detected by {\it wavdetect} have been removed. Each bin has a S/N$\geq$3, and an average photon flux as indicated by the color bar in units of photon cm$^{-2}$ s$^{-1}$ pixel$^{-1}$. Here 1 pixel corresponds to 0.98$\arcsec$. }
\label{fig:chandra_fig}
\end{figure}

We extracted and combined the 0.5--3 keV count images using {\it merge\_obs}. 
Background images were generated from the stowed observation files described in Section \ref{sec:data}.
We calculated the exposure maps using a weighted spectrum file generated by {\it make\_instmap\_weighted}, where the spectral model is an absorbed {\it apec} model with kT = 3 keV, which is the average temperature measured by previous {\it Chandra} studies \citep{Vikhlinin2005}.
After correcting the NXB-subtracted count image with the combined exposure map, we obtained the {\it Chandra} flux map.
A contour binning algorithm \citep{Sanders2006} was then applied to this flux map, to create a binned image with each bin reaching a signal-to-noise ratio of 3 (see Figure \ref{fig:chandra_fig}).

We employed the {\it wavdetect} algorithm on the combined counts images in the 0.5--3 keV band, adopting a false-positive probability threshold of $10^{-6}$ and wavelet scales of 1, 2, 4, 8, 16, 32 and 64 pixels.
A merged 90\% Enclosed Counts Fraction (ECF)  psfmap and merged exposure map generated by {\it merge\_obs} were supplied to the detection process. 
A total of 1175 point sources have been identified and were then excluded from the following analysis. 
Utilizing the same {\it Chandra} dataset, \citet{Shin2018} showed that the cumulative number count distribution (i.e., logN-logS curve) in Abell 133 is consistent with the CDF-S field \citet{Lehmer2012} within 1$\sigma$ uncertainty.
This indicates no explicit excess in the outskirts has been observed due to unidentified clumps among {\it wavdetect} sources.

\section{Surface brightness profile}
\label{sec:sb}
The azimuthally averaged surface brightness profiles of Abell 133 were extracted and fitted with the {\it pyproffit} package by \citet{Eckert2011}.
 In accordance with the analysis described in Section \ref{sec:clump},  both {\it Chandra} and {\it Suzaku} surface brightness profiles were fitted only in the outskirts, from 10.2$\arcmin$(0.7$r_{500}$) to 30$\arcmin$.
Before extraction, we first excluded the point sources above a photon flux threshold of 2$\times 10^{-6}$ photon cm$^{-2}$ s$^{-1}$, using the {\it Chandra} source list from {\it wavdetect} (see Section \ref{sec:chandra_imaging}).
We extracted concentric annuli centered on the X-ray peak of Abell 133 ( J2000 (RA, Dec) = (15.67333, -21.88011); \citealp{Randall2010} ) and binned the {\it Chandra} profile using a width of 2$\arcmin$ while for {\it Suzaku} a bin size of 3$\arcmin$ has been applied to account for its larger PSF size.

We fit the extracted surface brightness profiles with a power-law model ({\it pyproffit.PowerLaw}), $S_{x} = S_{0} (r / r_{0})^{-\alpha} + B$, where B represents the sky background.
Compared to the commonly used $\beta$-model, our best-fit power-law models gave a better description for the outskirts of Abell 133 since the $\beta$ parameter and core radius ($r_{0}$) of the $\beta$-model are strongly coupled outside the fitted radii \citep{Mohr1999}.

\begin{figure}
\centering
\includegraphics[width=0.49\textwidth]{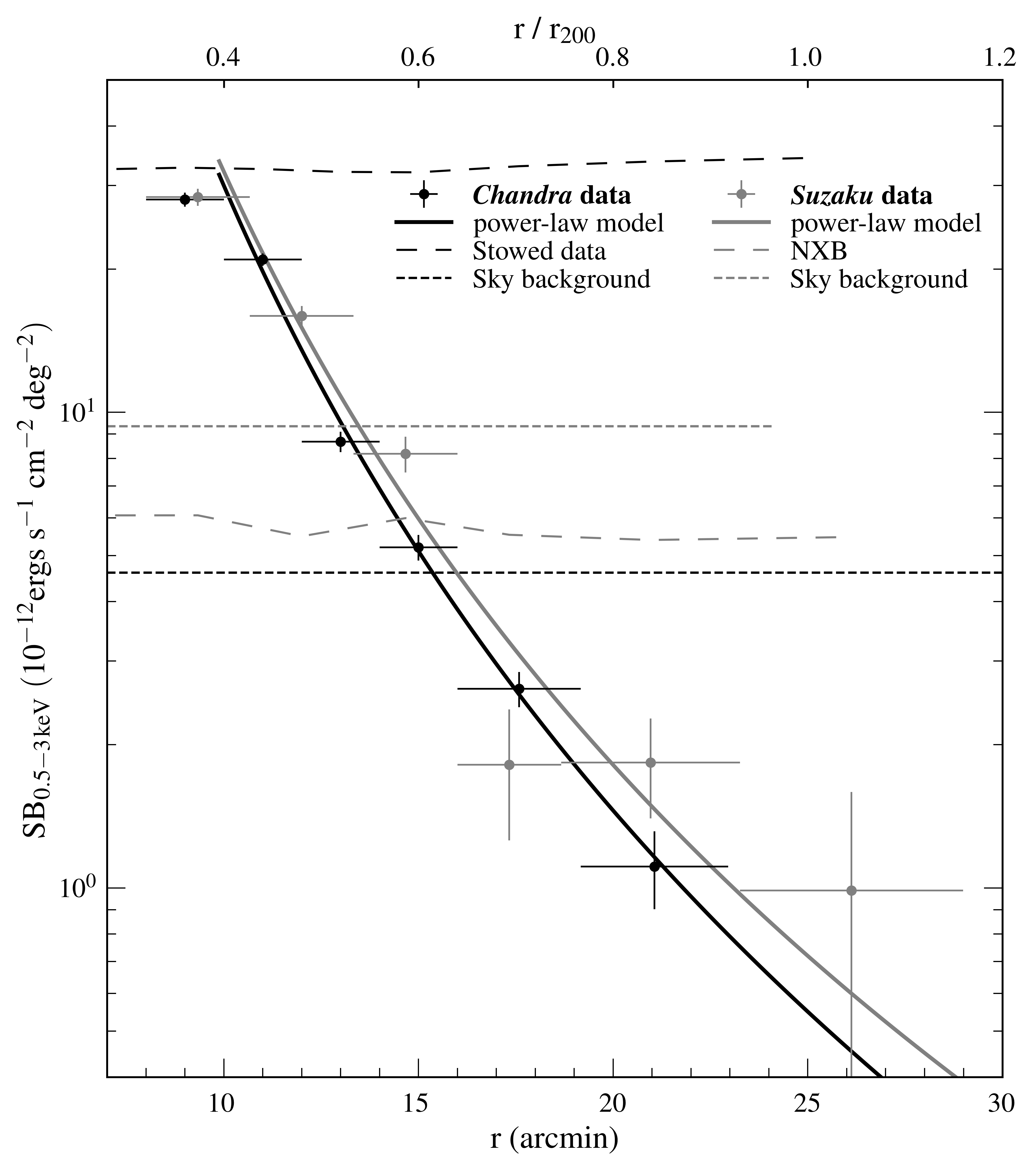}
\caption{0.5--3 keV background-subtracted azimuthally averaged surface brightness profile of the outskirts of Abell 133 obtained from {\it Chandra} (black) and {\it Suzaku} (grey) data with the corresponding best-fit power-law models overplotted. The {\it Suzaku} sky background level is relatively higher compared to {\it Chandra}, which reflects the excess emission from {\it Chandra}-selected point sources whose PSF is broader than the source extraction region.}
\label{fig:sb}
\end{figure}

In Figure \ref{fig:sb}, we show the background-subtracted surface brightness profiles with best-fit power-law models in units of erg s$^{-1}$ cm$^{-2}$ deg$^{-2}$.
The energy flux is converted from the photon flux with the same spectral model assumed in Section \ref{sec:chandra_imaging}.
The best-fit power-law indices are in good agreement, $\alpha = 4.37\pm0.24$ and $\alpha = 4.10\pm0.60$ for {\it Chandra} and {\it Suzaku} respectively, indicating the two profiles are well consistent.

\section{Clumping}
\label{sec:clump}
\citet{Zhuravleva2013} presented a useful method to characterize the ICM inhomogeneity.
Their work showed that the gas density in a given annulus roughly follows a log-normal distribution, while a high-density tail can be seen as a result of gas clumping.
In such a case, the mean and the median values of the distribution separate, as the median coincides with the peak of the log-normal profile and the mean shifts to a higher density.
As the X-ray emission is proportional to the square of the local density, it has been indicated by previous work that a similar distribution is also found for the surface brightness and flux \citep{Eckert2015,Mirakhor2021}.

\begin{figure*}
    \centering
    \includegraphics[width=0.45\textwidth]{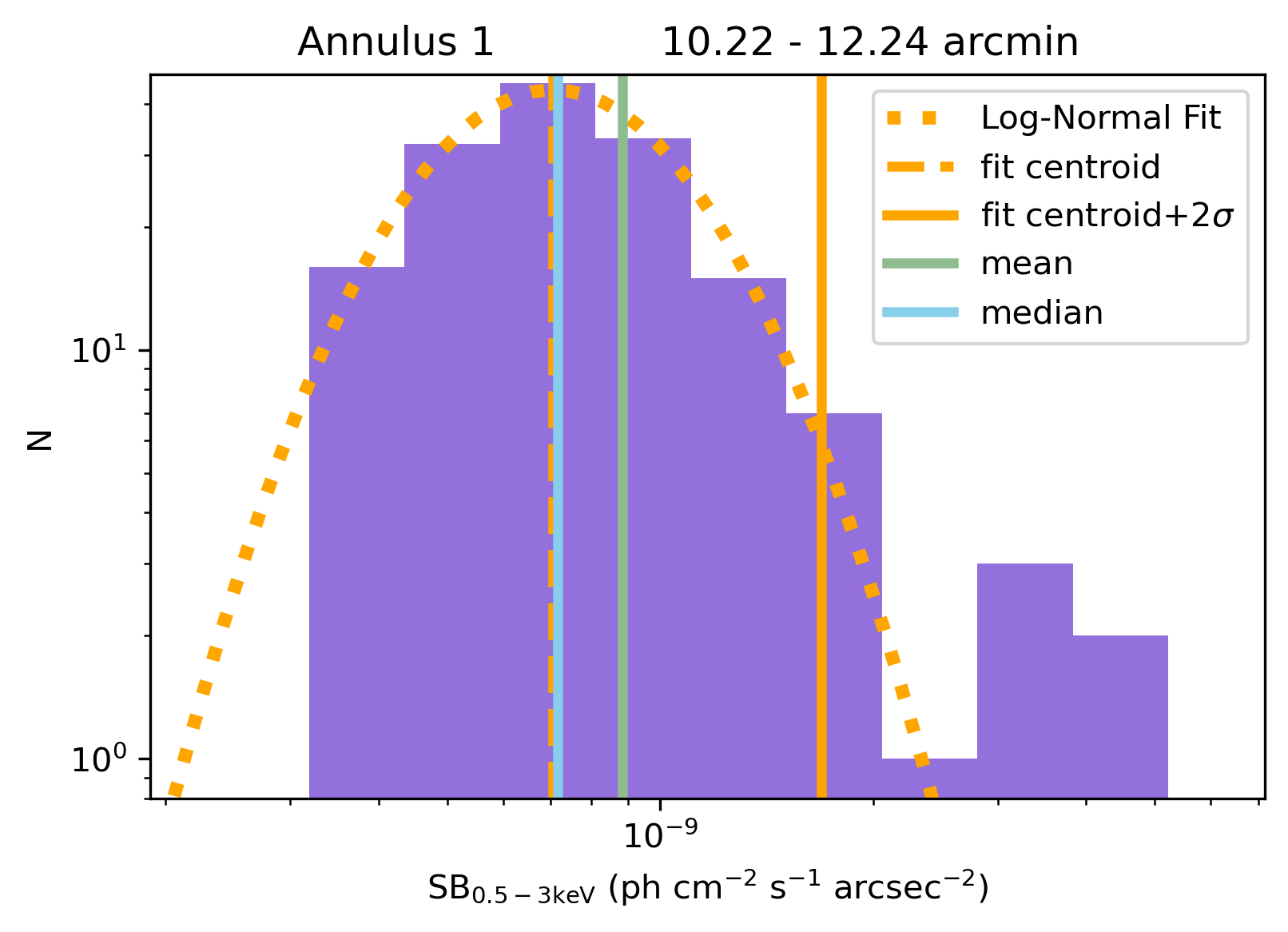}
    \includegraphics[width=0.45\textwidth]{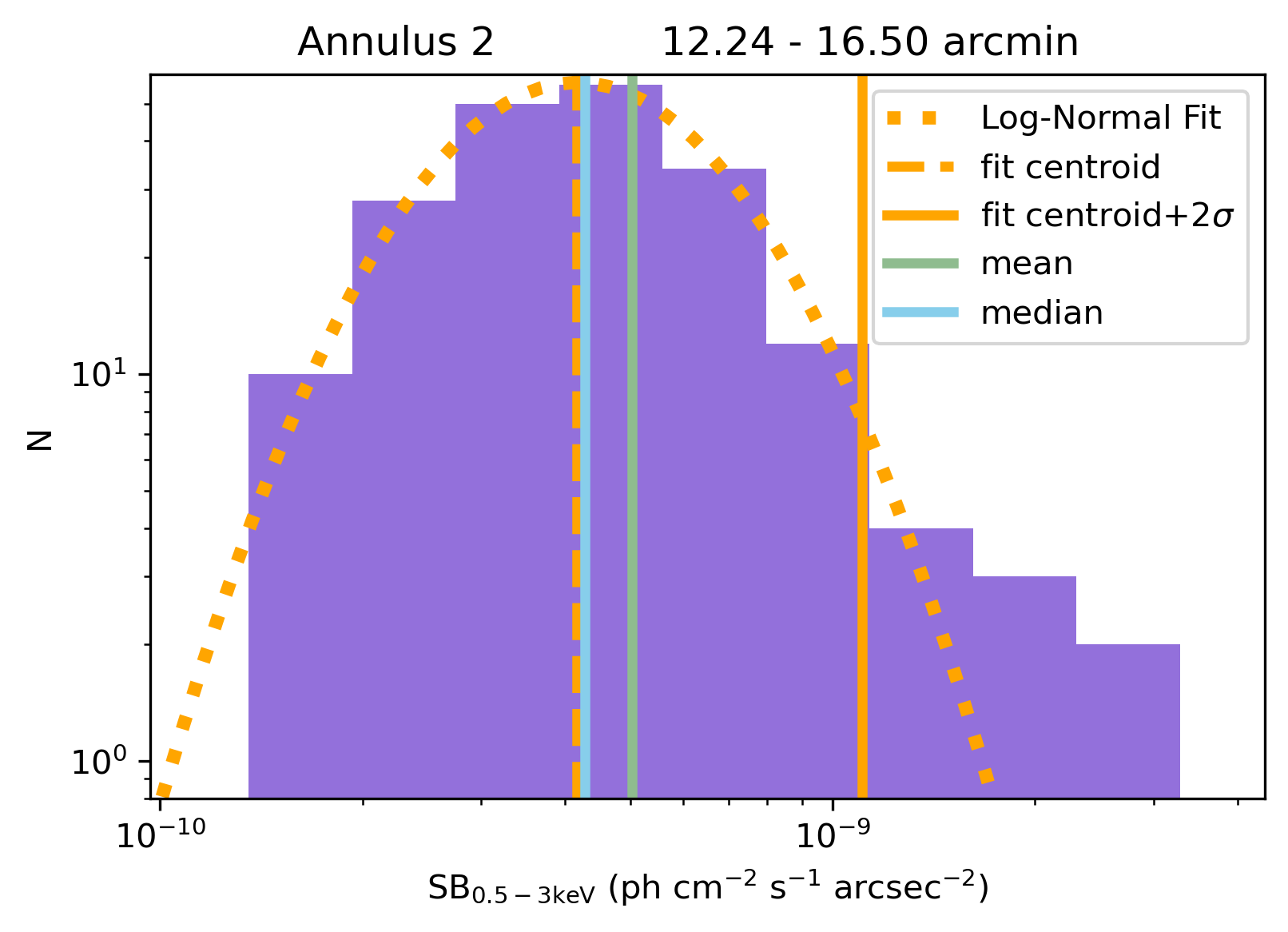}\\
      \includegraphics[width=0.45\textwidth]{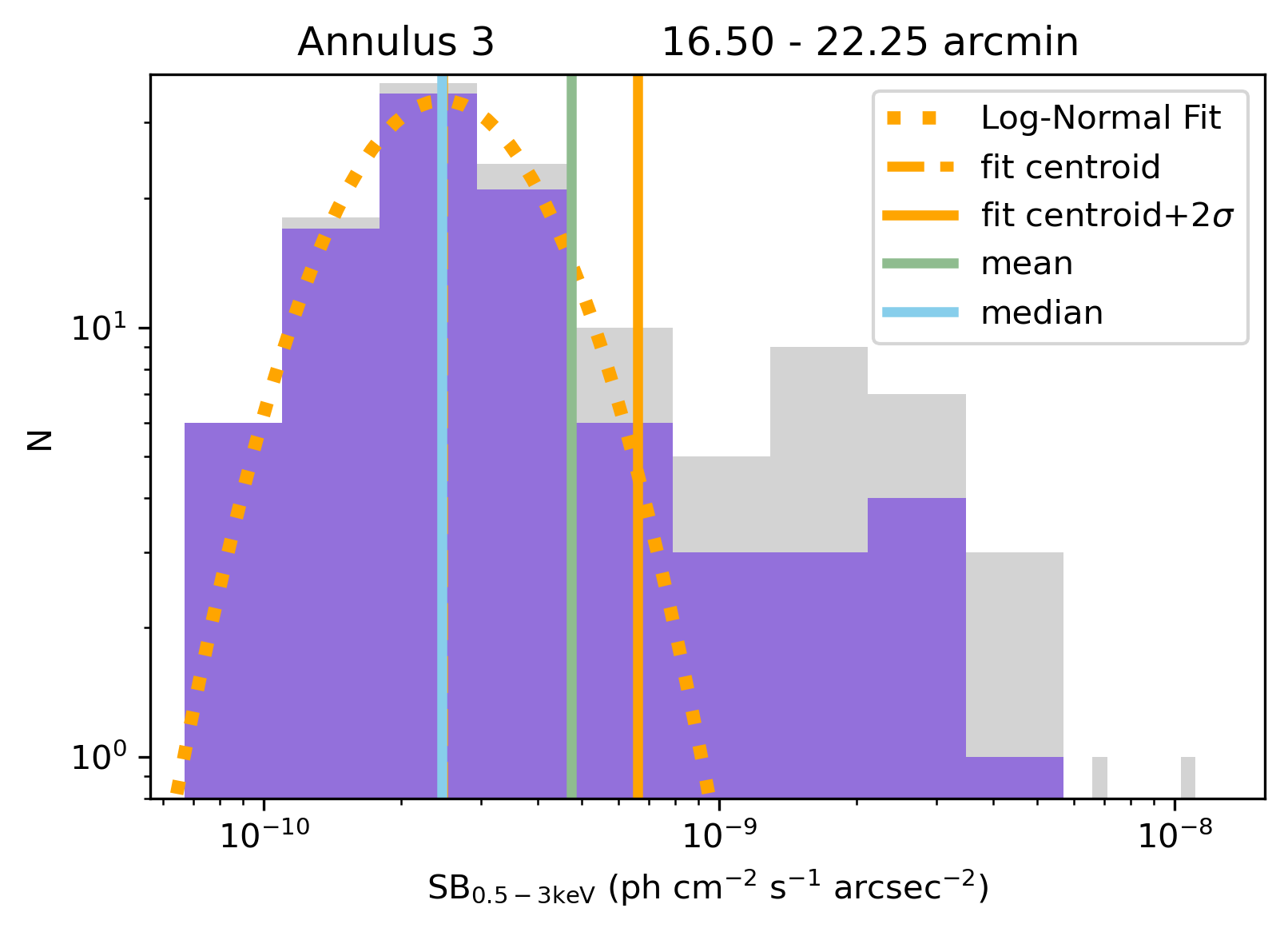}
        \includegraphics[width=0.45\textwidth]{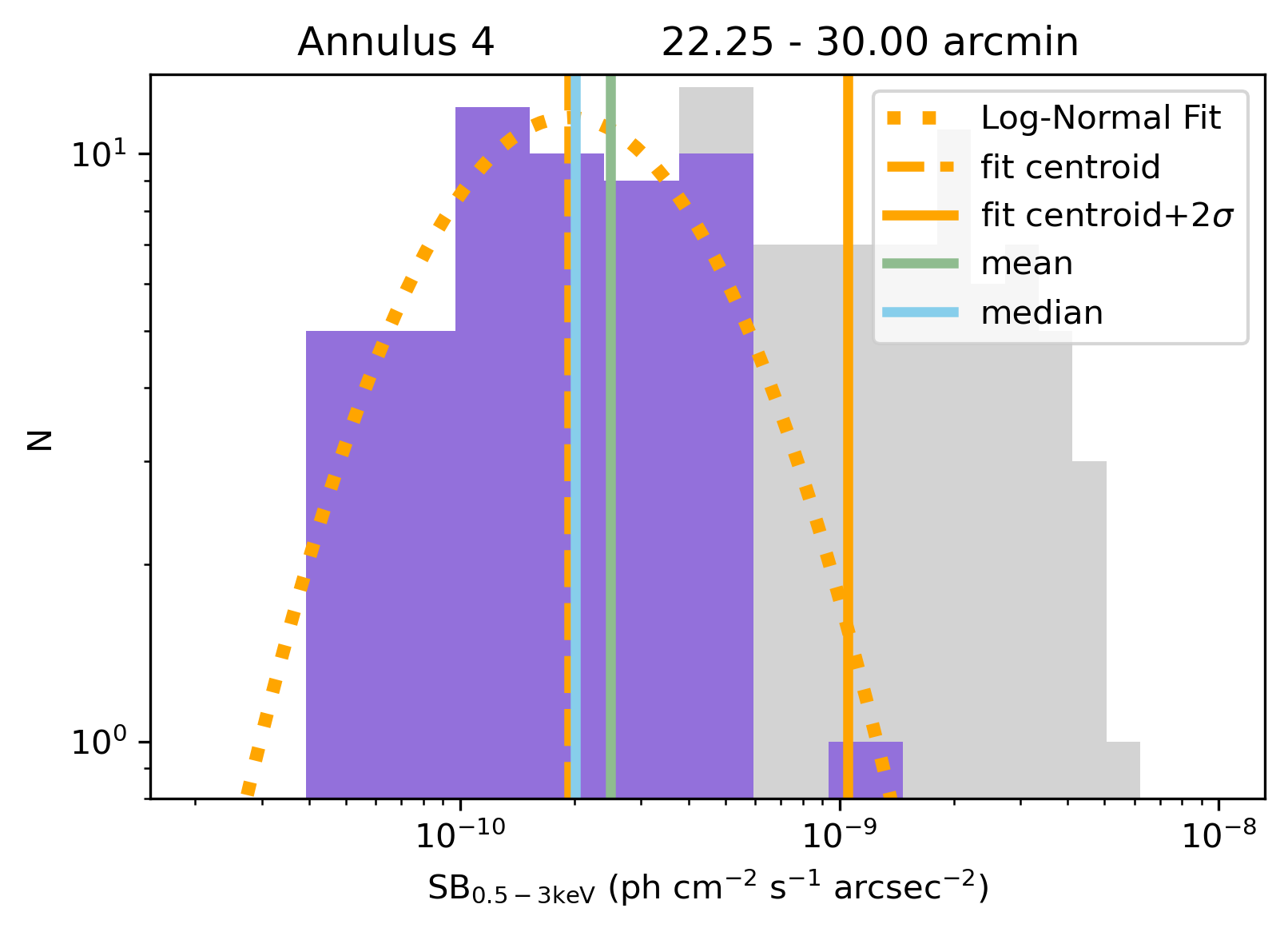}

    \caption{0.5-3 keV surface brightness distribution of Abell 133 outskirts obtained in 4 annuli, mapping from 10.2$\arcmin$ to 30.0$\arcmin$ as listed in Table \ref{tab:flux}. The histogram in grey shows the contribution from the large, bright diffuse features in the southeast, which has been excluded from further analysis. The solid orange line marks the 2$\sigma$ outlier threshold.}
    \label{fig:flux}
\end{figure*}

In the {\it Chandra} image, two large, bright clumps are immediately evident in the southeast (labelled BC1 and BC2). In order to search for additional clumps, we examine the surface brightness distribution of 4 annuli divided from 10.2$\arcmin$ (0.7$r_{500}$) to 30$\arcmin$ in Figure \ref{fig:flux}. 
The ICM of the inner core region within 0.7r$_{500}$ is affected by sloshing and merging events, and more importantly, we are mainly focused on the outskirts. Therefore we only search for clump candidates outside of 0.7r$_{500}$.
We fit each distribution with a log-normal profile and obtain the best-fit centroid and sigma. 
For each annulus, the median flux coincides well with the fit centroid, while the mean flux is shifted towards the high flux end, indicating an asymmetric distribution with a high-flux tail.
To quantify this asymmetry, we calculate the Fisher-Pearson coefficient of skewness of the distributions and estimate its 1-$\sigma$ error range using a bootstrap method with 10$^{5}$ resamplings.
The skewness measurements along with other surface brightness distribution parameters are listed in Table \ref{tab:flux}.
We notice that the skewness is positive in each annulus and roughly presents an increasing trend towards larger radii. 
We note that in the outermost annulus the presence of BC1 and BC2 would dominate and distort the formal log-normal profile (see Figure \ref{fig:flux}) and therefore these have been excluded from the calculation of the median and skewness.
We have selected 2$\sigma$ outliers from the surface brightness distribution of each annulus and then localized them on the adaptively-binned image.
We further merge neighbouring outlier bins and identify 16 clump candidates, which are listed in Table \ref{tab:clumps}. In case several neighbouring outlier bins were combined into a single clump, their individual significances were added in quadrature.
More details and tests for the X-ray clumping detection can be found in Kovács et al. 2023 (submitted). 

Combining the cluster catalog from the DESI Legacy Imaging Surveys \citep{Zou2021} and the r-band CFHT image taken as part of the Multi-Epoch Nearby Cluster Survey \citep{Sand2012,Graham2012}, we discuss the origin of individual clump candidates below.
\begin{itemize}
   \item \textbf{BC1}
    Overlaps with a large, bright background cluster (z$\sim$0.46). 
    \item \textbf{BC2}
    Overlaps with a large, bright background cluster (z$\sim$0.23). 
    \item \textbf{C3}
    The diffuse emission might be contributed from the nearby X-ray point source 2CXOJ010320.8-213317, although its X-ray luminosity is relatively low $S_{0.5-7}\le3.85\times10^{-15}$ erg s$^{-1}$ cm$^{-2}$.
    \item \textbf{C4}
    The candidate overlaps with a background cluster at the redshift of z$\sim$0.63.
    \item \textbf{C5}
    This candidate is adjacent to the X-ray point source 2CXO J010313.9-214103; however, its flux is weak ($S_{2-7}\le1.76\times10^{-15}$ erg s$^{-1}$ cm$^{-2}$), therefore the observed diffuse emission is not likely to originate from this point source. Within the contour, there is also a member galaxy (z$\sim$0.076+-0.027). However, the most likely origin of the surface brightness excess is an overdensity of galaxies with redshifts between 0.3 to 0.6, which are concentrated at the place where diffuse X-ray emission has been found.
    \item \textbf{C6}
    This candidate picks the emission from a weak point source 2CXO J010346.5-214653 ($S_{2-7}=2.50\times10^{-15}$ erg s$^{-1}$ cm$^{-2}$), which is missed by our point source detection.
    \item \textbf{C7}
    This is one of the nearest candidates to the core of Abell 133, which might hint to real ICM enhancement due to sloshing or feedback in the cluster core.
    \item \textbf{C8}
    Overlaps with a background cluster at the redshift of z$\sim$0.66.
    \item \textbf{C9}
    Located at the boundary of a background cluster (z$\sim$0.61). A concentration of background galaxies overlaps our candidate clump.
     \item \textbf{C10}
    A weak X-ray point source CXOGSG J010134.1-213943 (\citealp{Wang2016}), with a luminosity of $L_{0.3-8}\sim5.8\times10^{40}$ erg~s$^{-1}$.
    \item \textbf{C11}
    Overlaps with a background cluster (z$\sim$0.41). Although the brightest parts of the diffuse emission are removed as point sources, the remaining emission has been picked out by our method.
    \item \textbf{C12}
    Overlaps with a background cluster at the redshift of z$\sim$0.72.
    \item \textbf{C13} 
    This candidate close to the core region of Abell 133 could also be real ICM enhancement due to sloshing or feedback in the cluster core.
    \item \textbf{C14} 
    Same as C7 and C13.
    \item \textbf{C15}
    Overlaps with a galaxy WISEA J010204.45-220515.2. This candidate is identified as a very weak X-ray point source in \citet{Shin2018}, with flux of $S_{0.5-2}=2.70\times10^{-16}$ erg s$^{-1}$ cm$^{-2}$ .
    \item \textbf{C16}
    The X-ray peak of this candidate on the upper right corner, coincides with a galaxy WINGS J010146.15-220225.2 \citep{Varela2009}. 
   
\end{itemize}

We plot the cut-out images of {\it Chandra} and CFHT for each selected clump candidate in Figure \ref{fig:clumps}, using an image size of 0.2~Mpc $\times$ 0.2~Mpc, for better visual comparison.
The physical sizes of the clumps range from 8 kpc to $\sim$90 kpc in radius.
In conclusion, most of our clump candidates are background clusters or galaxies (BC1, BC2, C4, C5, C8, C9, C11, C12, C16), some are weak point sources missing from the source detection (C6, C10, C15), and the rest need further exploration (C3, C7, C13, C14).

We followed the definition of the {\it emissivity bias bx} in \citet{Eckert2015}, where $bx$ = SB$_{\rm mean}$ / SB$_{\rm median}$, to check the difference before and after removal of our identified clumps.
For each annulus, we used the mean and median surface brightness of bins within this annulus for SB$_{\rm mean}$ and SB$_{\rm median}$.
In Figure \ref{fig:bx}, we plot the azimuthally averaged emissivity bias profiles with and without the resolved clumps, where we see a difference beyond $r_{500}$.
We emphasize that the difference in the outermost annulus is underestimated since two bright clumps (BC1 and BC2) have been excluded before the clumping analysis.
We further compare the measurements of Abell 133 with the average emissivity bias profile measured from 31 clusters in the redshift range 0.04–0.2 observed with the ROSAT/Position
Sensitive Proportional Counter (PSPC) \citep{Eckert2015}. 
The two profiles match well within $r_{200}$, with $bx$ being lower -- as expected -- after the removal of the resolved clumps.
The clump-excluded profile (orange points in Figure \ref{fig:bx}) also implies that unresolved clumping remains.
However, it is worth noting that since the sky background is included in the surface brightness measurements, the emissivity bias can not quantitatively trace the clumping effect of the real ICM, and thus cannot be used to correct the unresolved clumping.

\begin{table*}
\begin{center}
\begin{small}
\caption{Surface brightness distribution parameters.}
\label{tab:flux}
\begin{tabular}{l c c c c c c c c c c c c}
\hline\hline
ID & r$_{in}$  & r$_{out}$ & N$_{bins}$ & N$_{clumps}$ & SB$_{\rm mean}$  & SB$_{\rm median}$ & Skewness & $b_{x}$ & SB$_{\rm mean}^{\star}$  & SB$_{\rm median}^{\star}$ &  Skewness$^{\star}$ & $b_{x}^{\star}$ \\
(1) & (2) & (3) & (4) & (5) & (6) & (7) & (8) & (9) & (10) & (11) & (12) & (13) \\\hline
 1 & 10.2 & 12.2& 155 & 4 & 8.85$^{+0.61}_{-0.45}$ &7.16$^{+0.34}_{-0.33}$ &3.72$\pm$0.60 & 1.24$\pm$0.08 & $7.56^{+0.25}_{-0.23}$ &$6.96^{+0.3}_{-0.27}$ &0.91$\pm$0.16 & 1.09$\pm$0.05\\
 2 & 12.2  & 16.5& 199 & 5 & 5.04$^{+0.30}_{-0.24}$&4.28$_{-0.21}^{+0.12}$ &3.79$\pm$0.66 & 1.18$\pm$0.08 & 4.39$^{+0.14}_{-0.13}$ &4.19$^{+0.11}_{-0.29}$ &0.94$\pm$0.13 & 1.05$\pm$0.08\\
 3 & 16.5 &22.3 &97 & 5 &4.74$^{+0.96}_{-0.64}$  &2.47$^{+0.10}_{-0.34}$ & 4.32$\pm$0.85& 1.92$\pm$0.36 & $2.51^{+0.14}_{-0.12}$ &2.12$^{+0.24}_{-0.08}$ & 1.04$\pm$0.24 & 1.18$\pm$0.07\\
 4 & 22.3 & 30.0 & 53& 2 & 2.49$^{+0.38}_{-0.25}$& 2.02$^{+0.10}_{-0.62}$ & 3.03$\pm$1.16& 1.23$\pm$0.40 & 2.25$^{+0.22}_{-0.19}$ &1.94$^{+0.10}_{-0.51}$ &0.67 $\pm$0.23 & 1.16$\pm$0.32\\

\hline
\end{tabular}
\end{small}
\end{center}
\tablefoot{Columns are as follows. (1) Annulus ID; (2)(3) inner and outer radius, in arcmin; (4) number of bins in each annulus (note that a total of 504 of bins are included in the annuli); (5) number of $2\sigma$ outliers after combining neighbouring bins; (6)(7) mean and median surface brightness in 0.5--3 keV, in units of $10^{-10}$ photon cm$^{-2}$ s$^{-1}$ arcsec$^{-2}$. (8) Fisher-Pearson coefficient of skewness; (9) Emissivity bias defined as $b_{x}$ = SB$_{\rm mean}$ / SB$_{\rm median}$. (10)(11)(12)(13) are measurements with the 16 clump candidates removed.}
\end{table*}

\begin{table*}
\begin{center}
\caption{Clump candidates.}
\label{tab:clumps}
\begin{tabular}{l  c c c c c c c c c}
\hline\hline
Clump ID & R.A. & Dec  & $N_{bins}$ & F$_{0.5-3 \rm keV}$  & Area & Significance  \\
(1) & (2) & (3) & (4) & (5) & (6) &(7) \\\hline
BC1 &  1:03:34.1026 &     -22:12:58.368 & 48&    3.37$\times10^{-5}$ &  17419 &      21.1 \\
BC2 &  1:03:12.3089 &    -22:12:49.189 & 13 &      1.58$\times10^{-5}$  &  8188 &       18.0 \\
C3 &  1:03:19.7836 & -21:33:26.717&  3 &  1.73$\times10^{-6}$ &   824 &      7.62 \\
C4 &  1:03:08.8609 &    -21:38:51.309 & 1&   9.24$\times10^{-7}$ &      696 &    2.37 \\
C5 &  1:03:13.4436 &    -21:40:53.937 & 3 &    3.33$\times10^{-6}$ &    1862 &       5.24 \\
C6 &  1:03:46.7412 &      -21:46:45.169 & 1&    3.48$\times10^{-7}$ &    184 &         3.09 \\
C7 &  1:03:23.0879 &      -21:48:28.149 &1 &   8.32$\times10^{-7}$ &  212 &     3.93 \\
C8 &  1:03:58.0752 &       -21:52:57.118 & 6 &    4.38$\times10^{-6}$ &    4290 &     8.33 \\
C9 & 1:03:33.8031 &    -21:53:59.633 & 7 &     4.38$\times10^{-6}$ &     1446 &        9.82 \\
C10 & 1:01:34.3747 &        -21:39:44.730 & 2&      1.14$\times10^{-6}$ &   308 &     8.06 \\
C11 & 1:01:58.5747 &       -21:27:18.801 & 1 &      1.72$\times10^{-6}$ &    1180 &       2.39 \\
C12 & 1:04:14.2142 & -21:48:07.787  & 1   & 4.78$\times10^{-7}$ & 401 & 3.19 \\
C13 & 1:02:11.6771 & -22:00:34.794  & 1   & 3.53$\times10^{-6}$ & 2024 & 2.07 \\
C14 & 1:02:06.3268 & -22:00:47.524  & 1  & 1.22$\times10^{-6}$ & 664 & 2.18 \\
C15 & 1:02:03.7572 & -22:05:09.921  & 1 &  6.34$\times10^{-7}$ & 525 & 2.17 \\
C16 & 1:01:47.1879 & -22:02:33.340  & 1 & 8.83$\times10^{-7}$ & 738 & 2.16 \\
\hline
\end{tabular}
\end{center}
\tablefoot{Columns are as follows. (1) Clump ID; (2)(3) central coordinates of each clump; (4) the number of neighboring outlier bins that were grouped; (5) total flux of each clump candidate, in units of photon cm$^{-2}$ s$^{-1}$; (6) area of the clump, in units of arcsec$^{2}$; (7) significance of the clump.}
\end{table*}

\begin{figure*}
    \includegraphics[trim=2cm 2cm 2cm 0cm, clip, width=.245\textwidth]{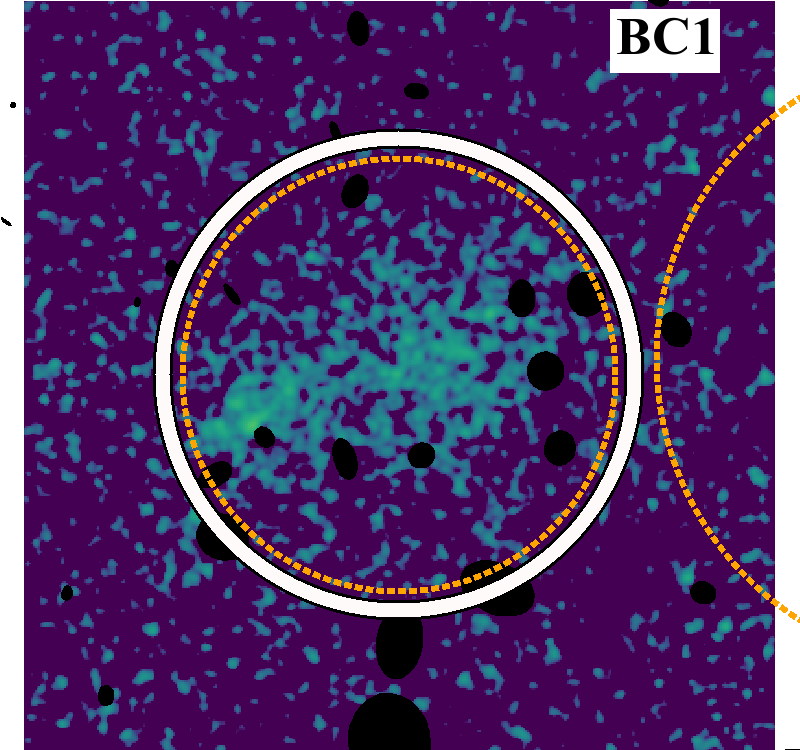}
     \includegraphics[trim=2cm 2cm 2cm 0cm, clip, width=.245\textwidth]{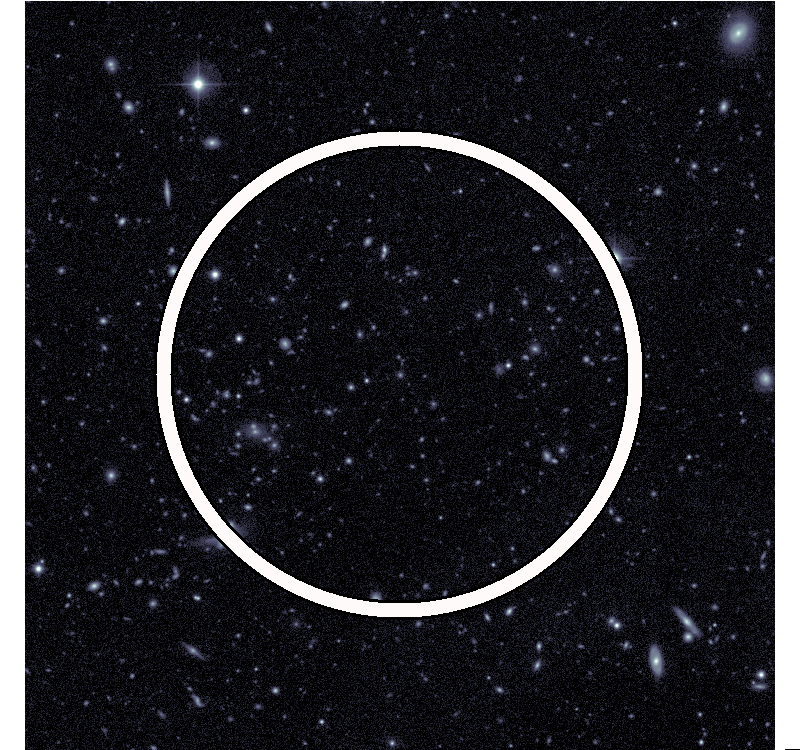}
     \includegraphics[trim=2cm 2cm 2cm 0cm, clip, width=.245\textwidth]{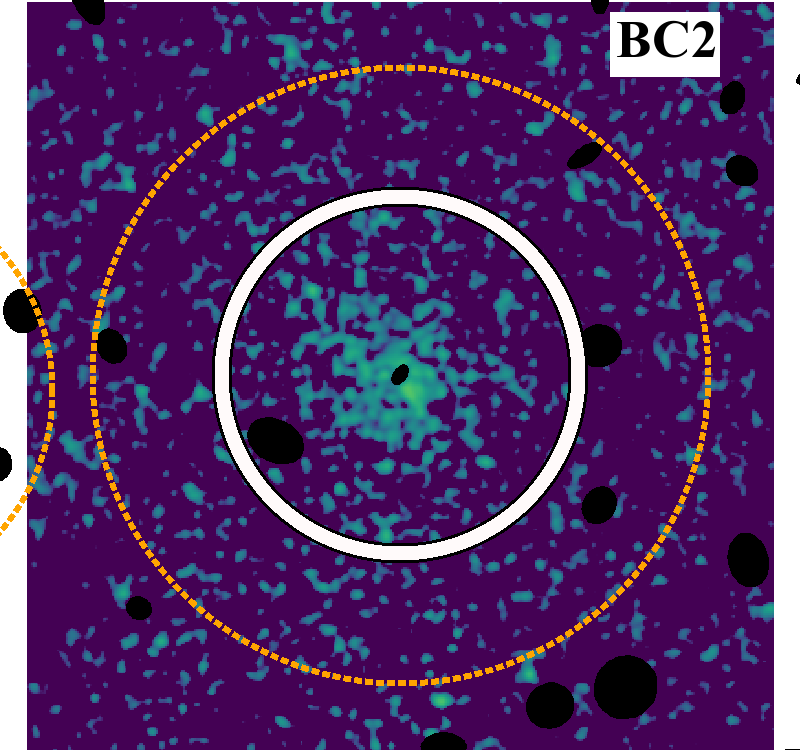}
     \includegraphics[trim=2cm 2cm 2cm 0cm, clip, width=.245\textwidth]{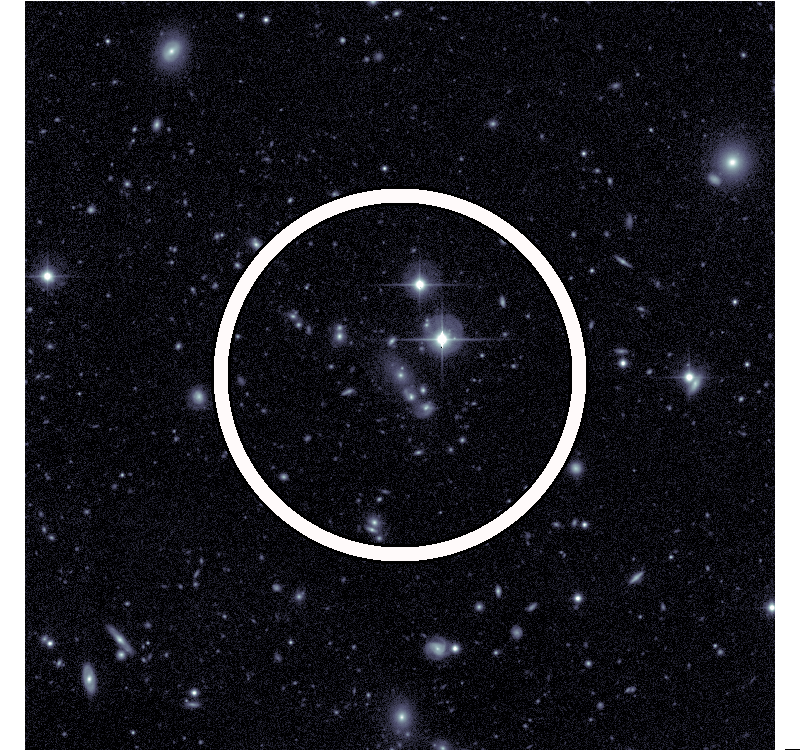}
      \hspace{.25cm} 
    \includegraphics[trim=2cm 2cm 2cm 0cm, clip, width=.245\textwidth]{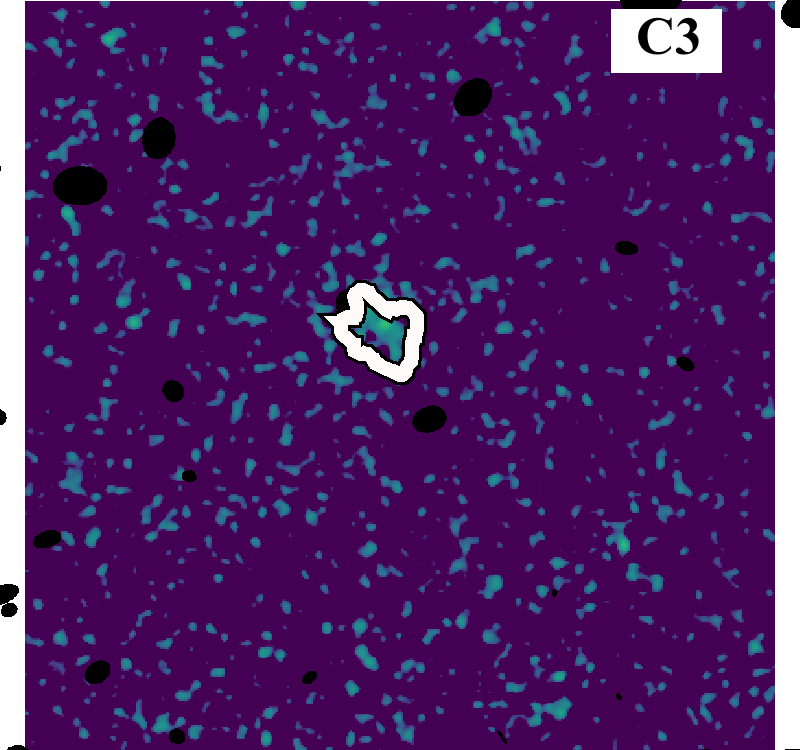}
     \includegraphics[trim=2cm 2cm 2cm 0cm, clip, width=.245\textwidth]{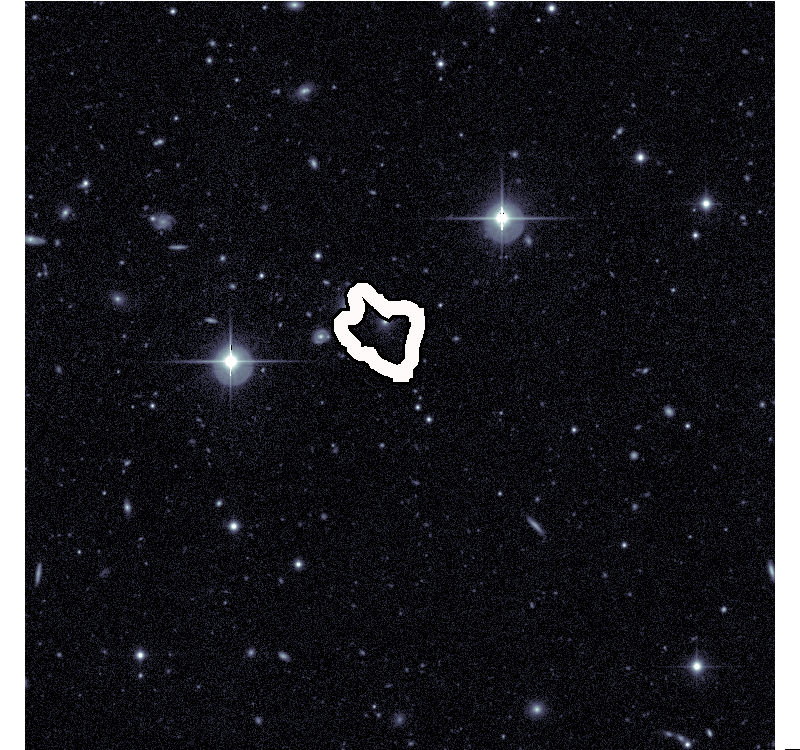}
     \includegraphics[trim=2cm 2cm 2cm 0cm, clip, width=.245\textwidth]{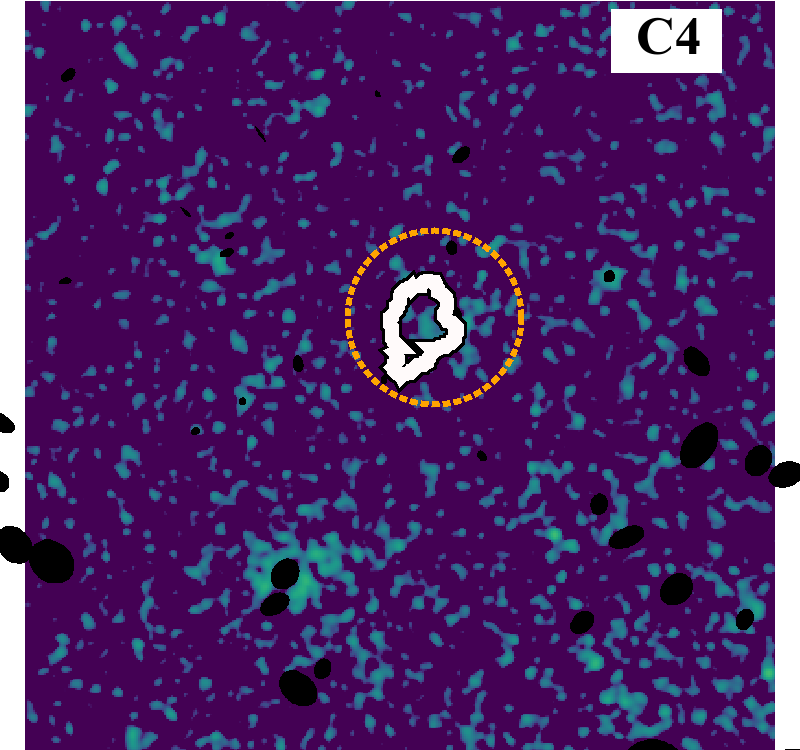}
     \includegraphics[trim=2cm 2cm 2cm 0cm, clip, width=.245\textwidth]{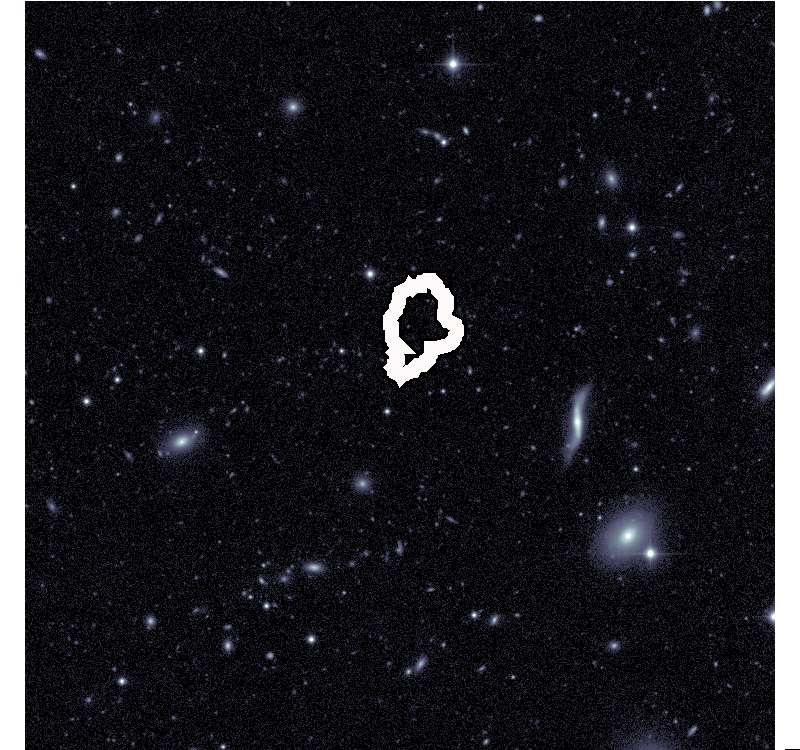}
      \hspace{.25cm} 
     \includegraphics[trim=2cm 2cm 2cm 0cm, clip, width=.245\textwidth]{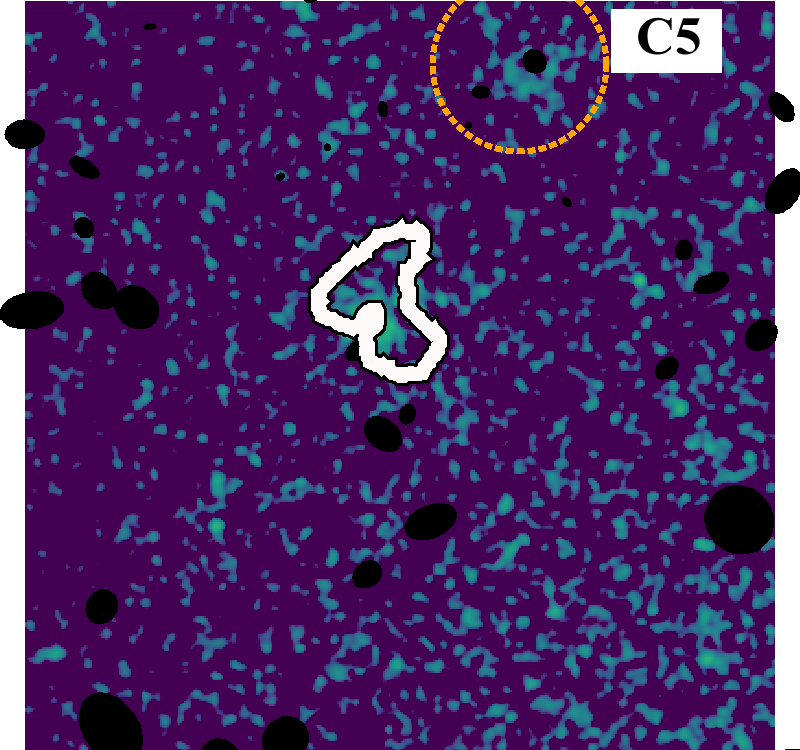}
     \includegraphics[trim=2cm 2cm 2cm 0cm, clip, width=.245\textwidth]{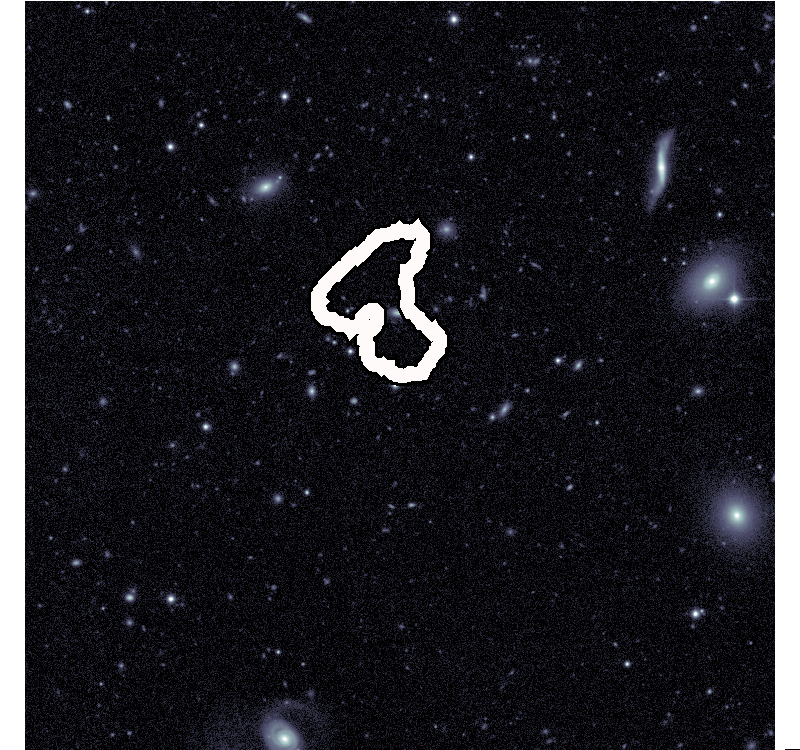}
     \includegraphics[trim=2cm 2cm 2cm 0cm, clip, width=.245\textwidth]{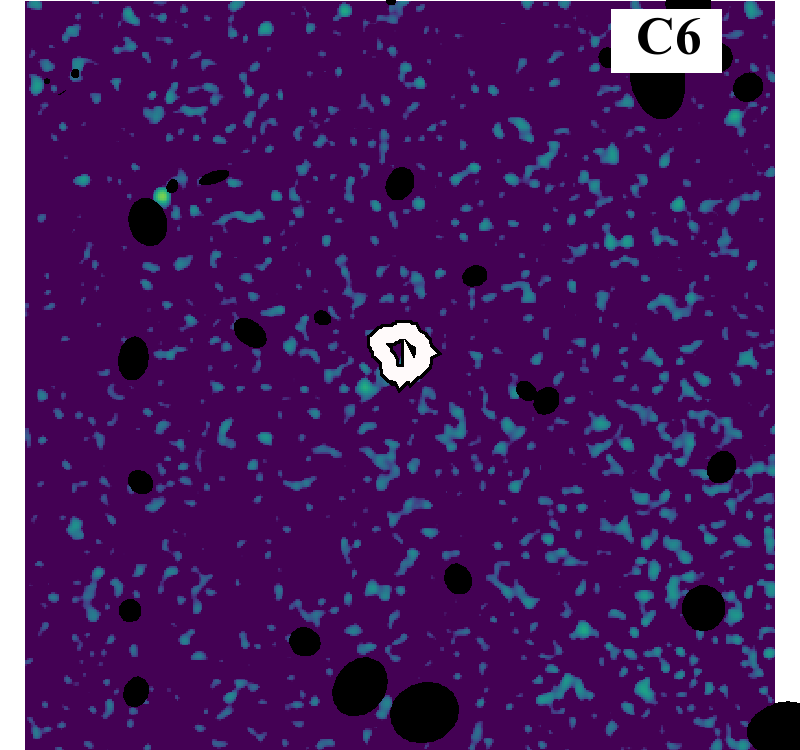}
     \includegraphics[trim=2cm 2cm 2cm 0cm, clip, width=.245\textwidth]{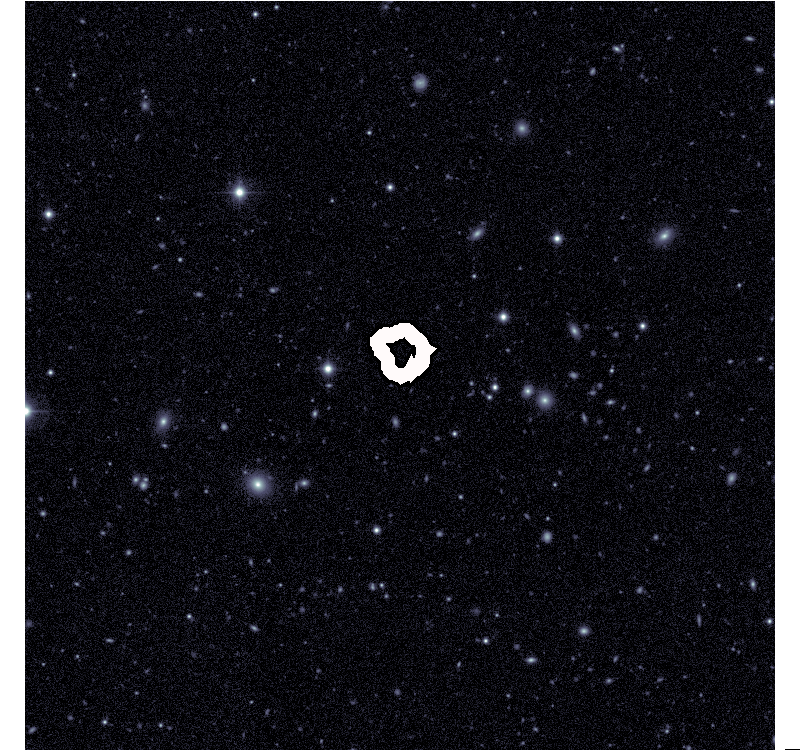}
      \hspace{.25cm} 
    \includegraphics[trim=2cm 2cm 2cm 0cm, clip, width=.245\textwidth]{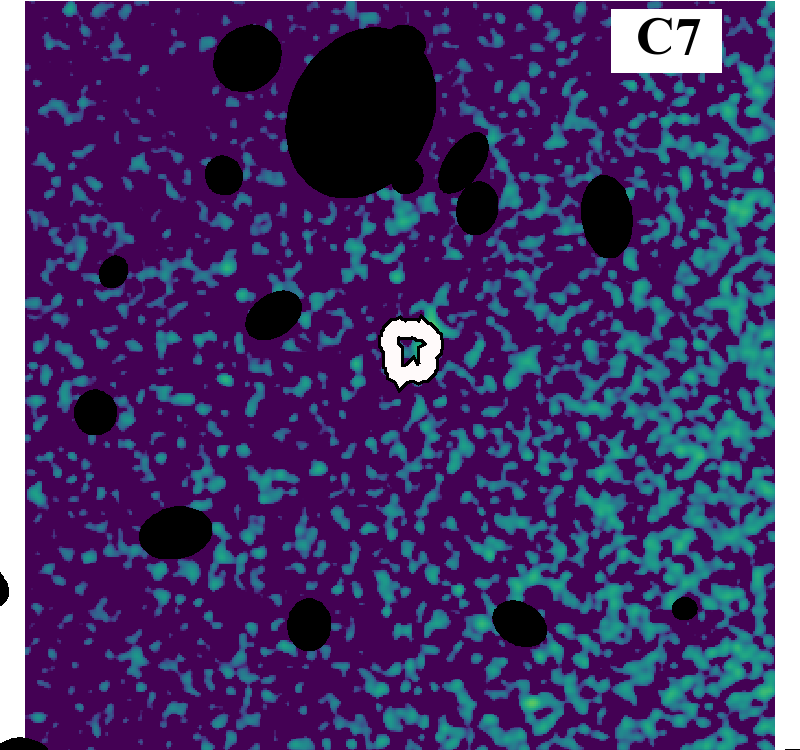}
     \includegraphics[trim=2cm 2cm 2cm 0cm, clip, width=.245\textwidth]{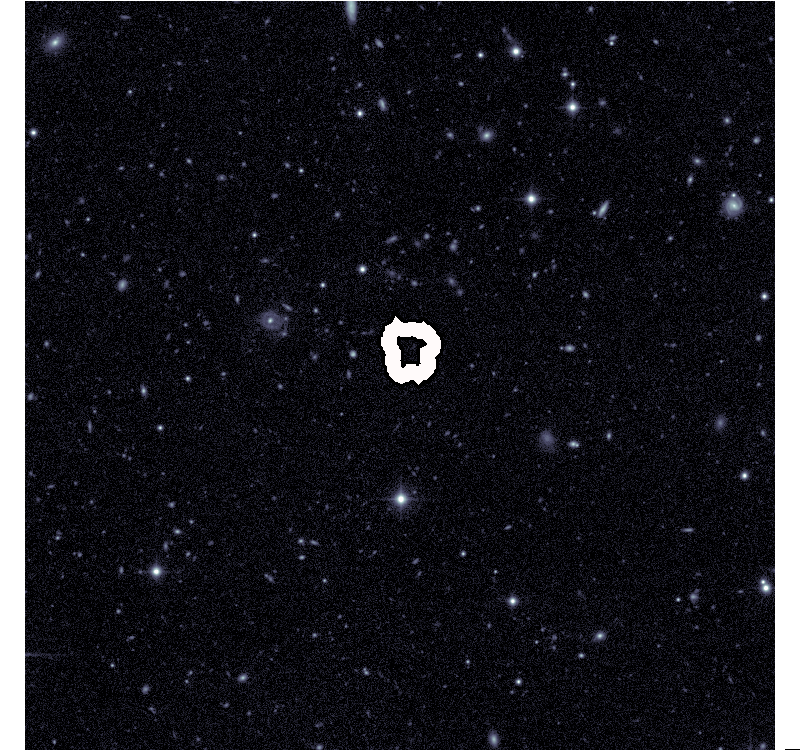}
     \includegraphics[trim=2cm 2cm 2cm 0cm, clip, width=.245\textwidth]{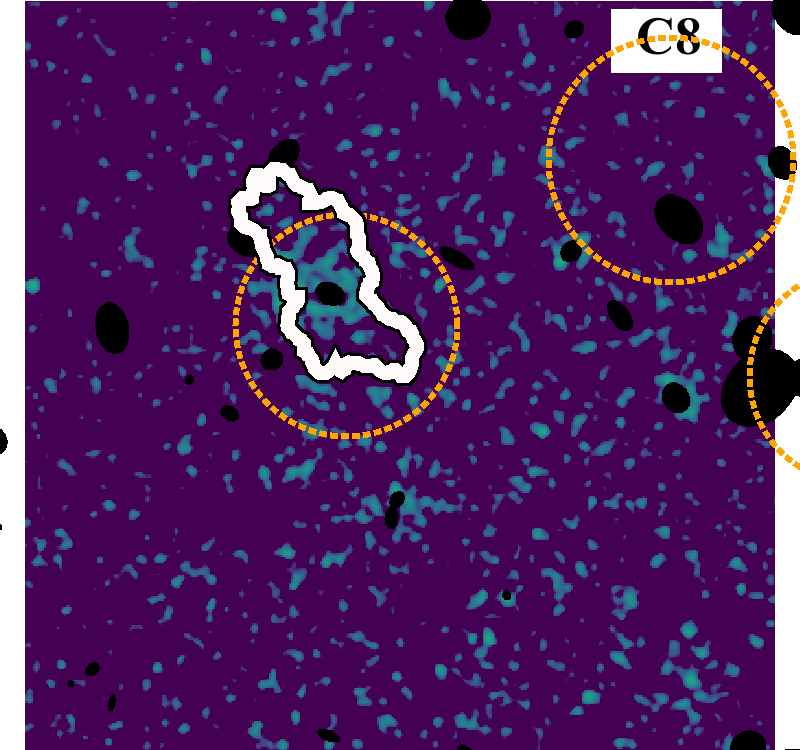}
     \includegraphics[trim=2cm 2cm 2cm 0cm, clip, width=.245\textwidth]{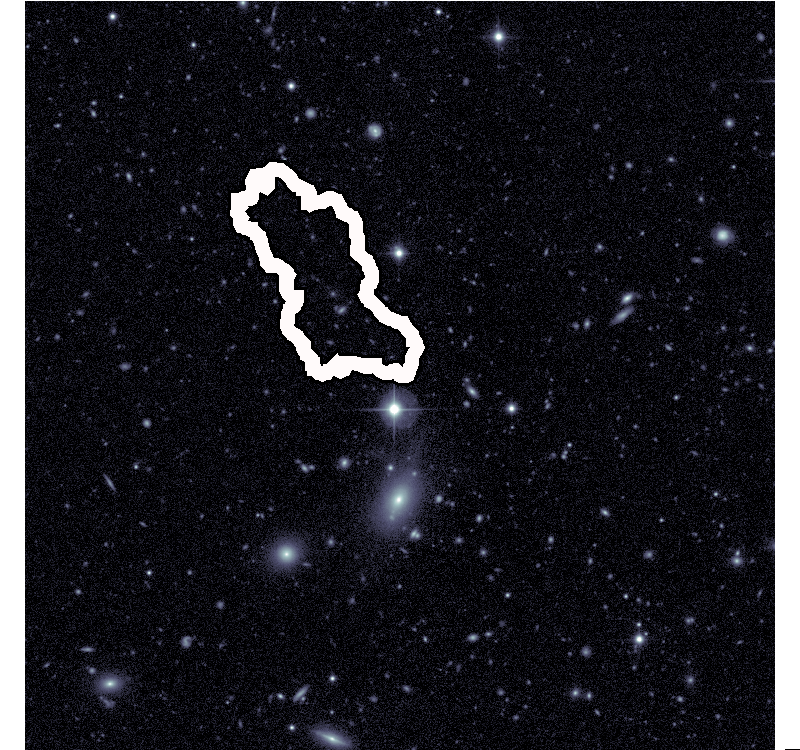}    
     \caption{0.5-3 keV cut-out images for clump candidates (white regions). {\it Left:} Each {\it Chandra} image has a size of $200 \arcsec \times 200 \arcsec$ (0.2 Mpc $\times$ 0.2 Mpc) and is smoothed with a Gaussian of radius $6$. Filled black ellipses represent excluded wavdetect sources. The orange dashed circles mark the $r_{500}$ radius of background clusters \citep{Zou2021}. {\it Right:} r-band CFHT image showing the same field of view. }

    \label{fig:clumps}
\end{figure*}

\begin{figure*}
    \ContinuedFloat
    \includegraphics[trim=2cm 2cm 2cm 0cm, clip, width=.245\textwidth]{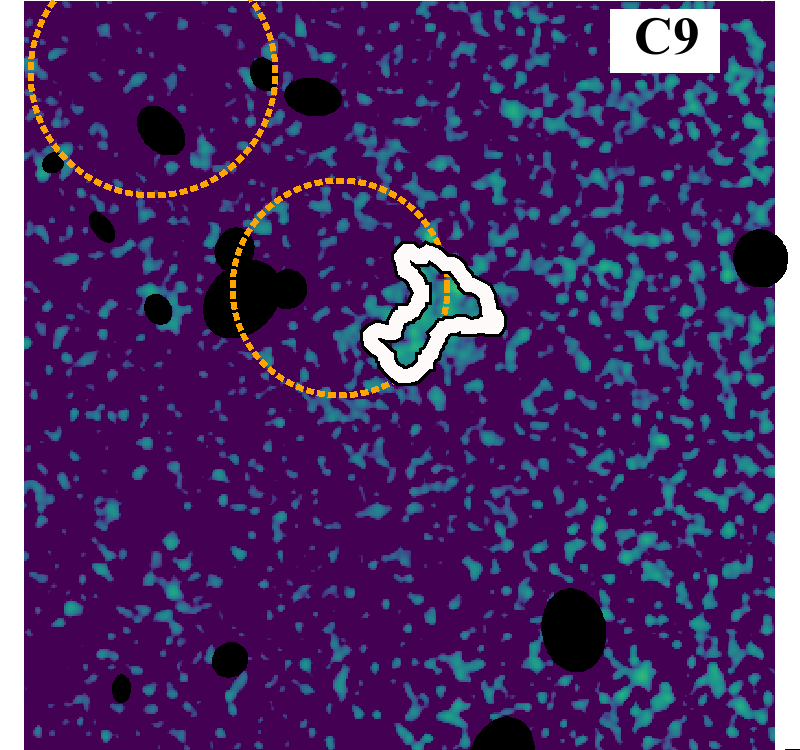}
     \includegraphics[trim=2cm 2cm 2cm 0cm, clip, width=.245\textwidth]{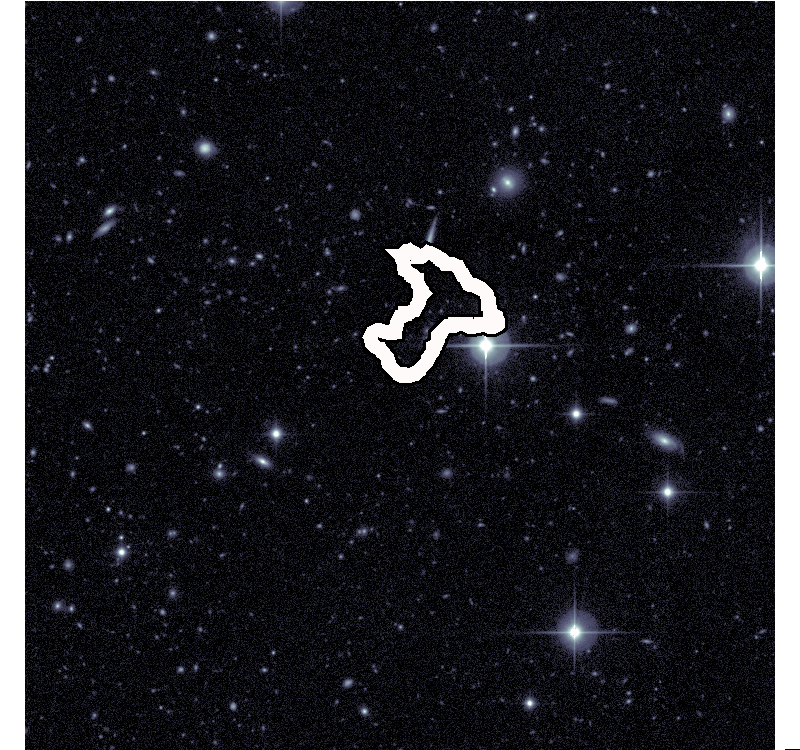}
     \includegraphics[trim=2cm 2cm 2cm 0cm, clip, width=.245\textwidth]{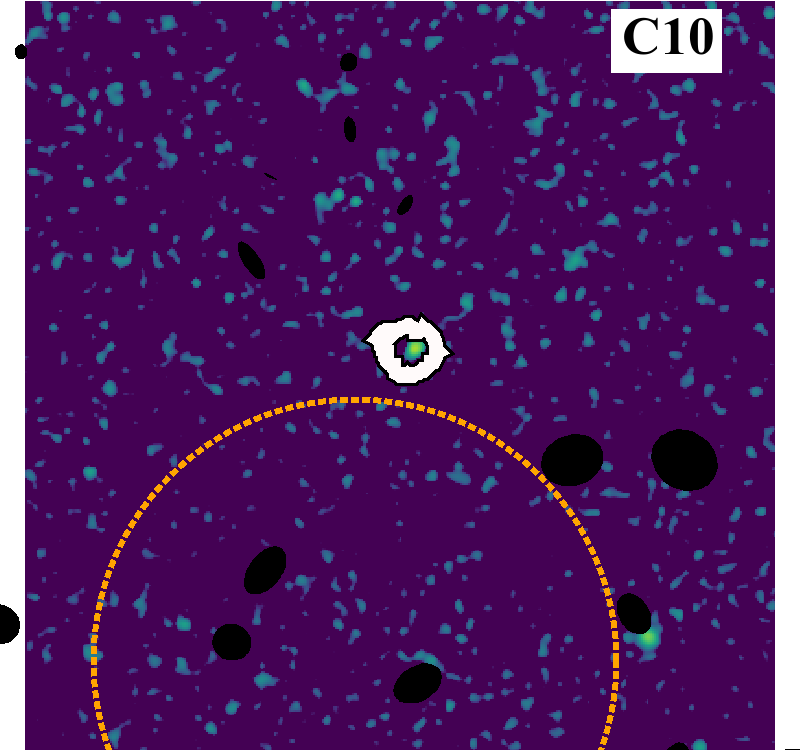}
     \includegraphics[trim=2cm 2cm 2cm 0cm, clip, width=.245\textwidth]{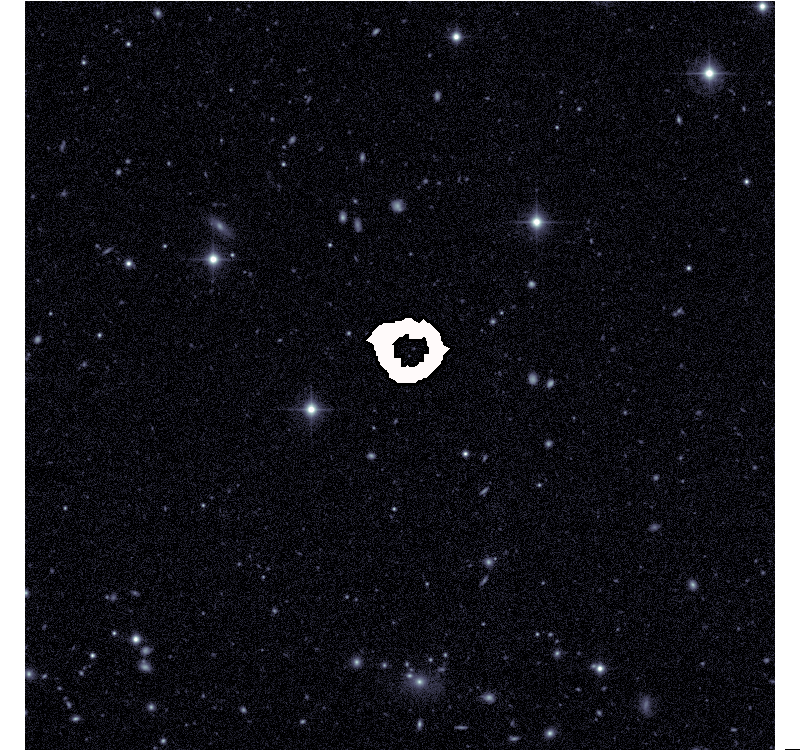}
      \hspace{.25cm} 
    \includegraphics[trim=2cm 2cm 2cm 0cm, clip, width=.245\textwidth]{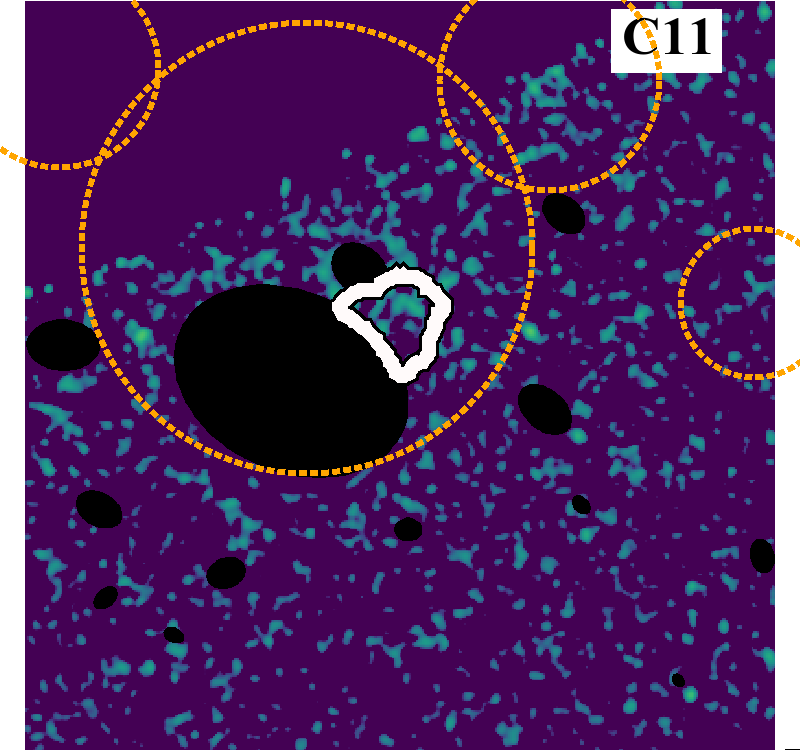}
     \includegraphics[trim=2cm 2cm 2cm 0cm, clip, width=.245\textwidth]{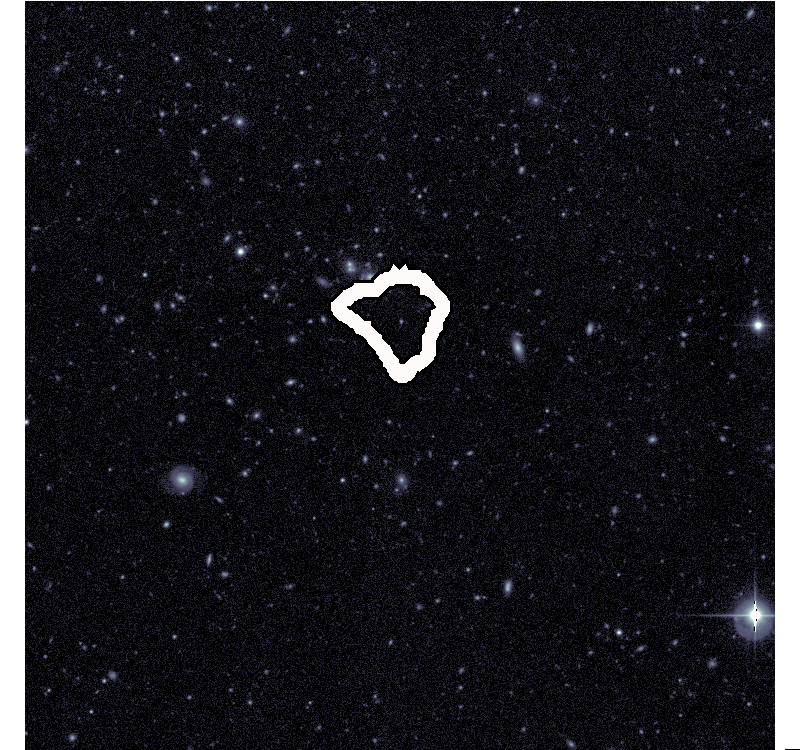}
     \includegraphics[trim=2cm 2cm 2cm 0cm, clip, width=.245\textwidth]{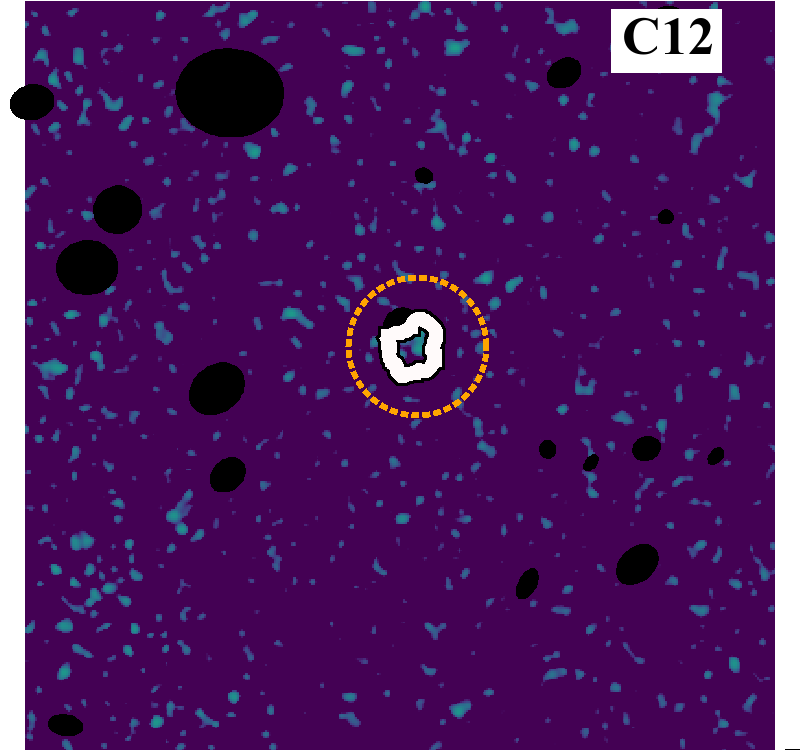}
     \includegraphics[trim=2cm 2cm 2cm 0cm, clip, width=.245\textwidth]{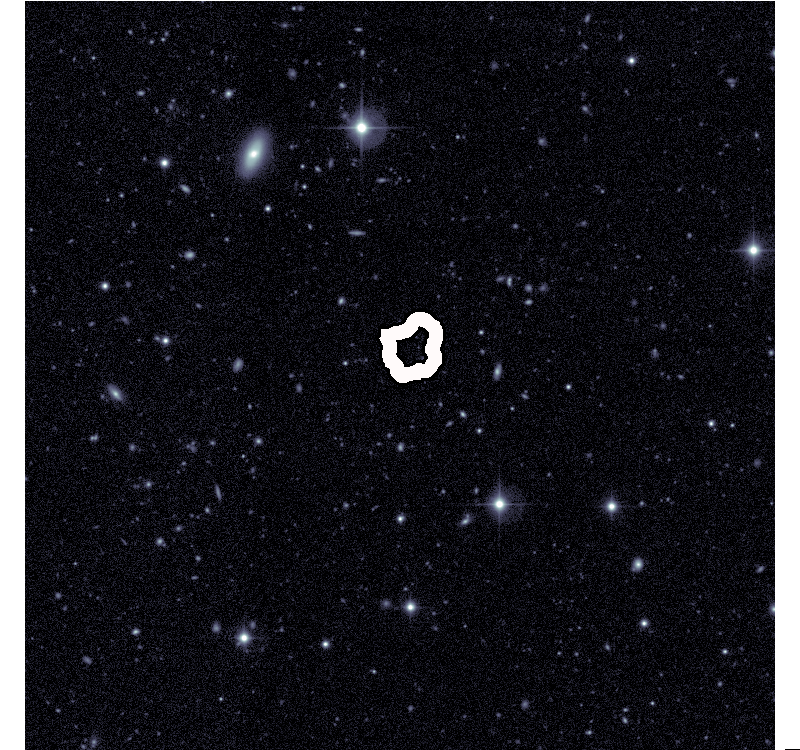}
      \hspace{.25cm} 
     \includegraphics[trim=2cm 2cm 2cm 0cm, clip, width=.245\textwidth]{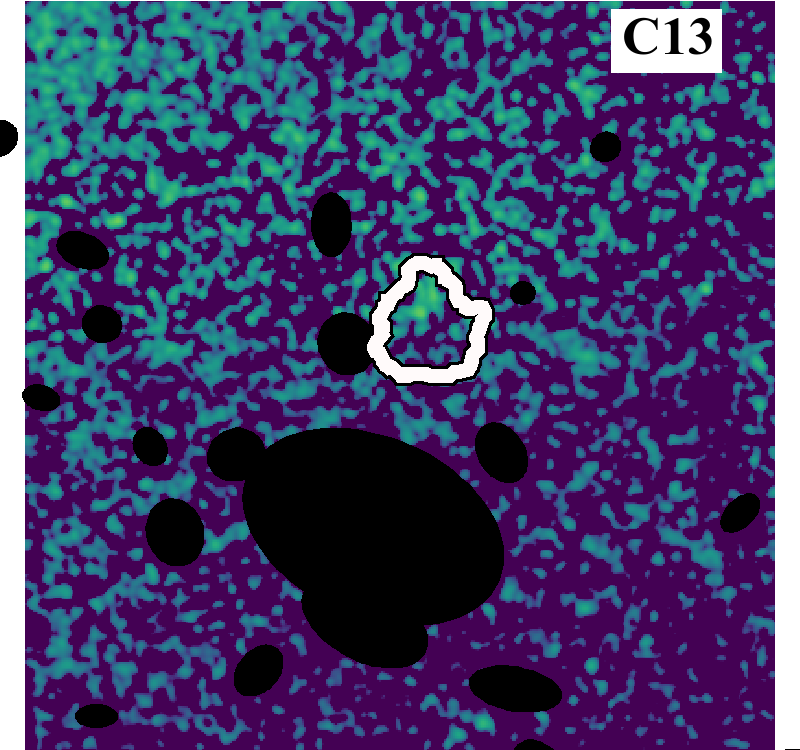}
     \includegraphics[trim=2cm 2cm 2cm 0cm, clip, width=.245\textwidth]{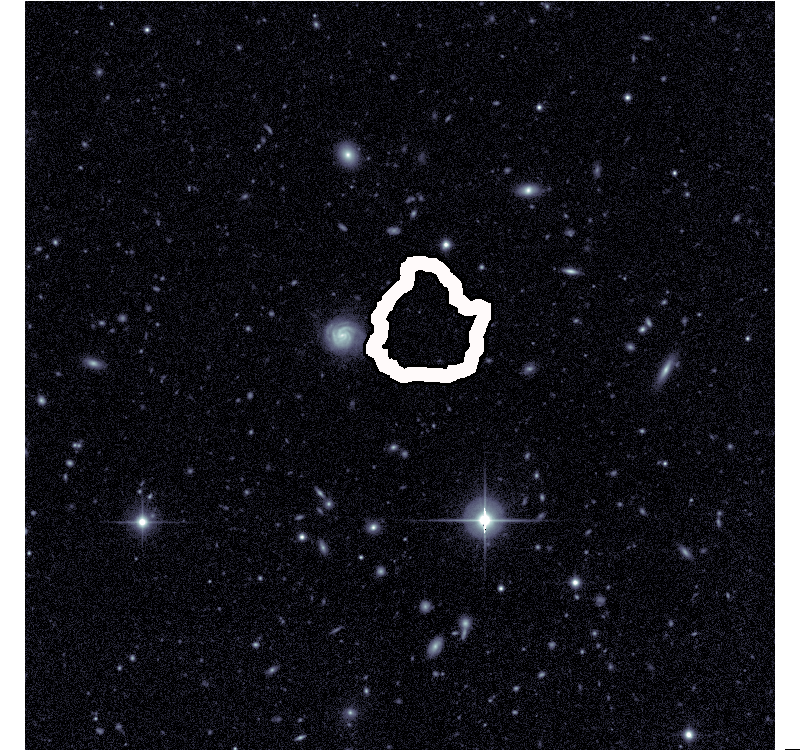}
     \includegraphics[trim=2cm 2cm 2cm 0cm, clip, width=.245\textwidth]{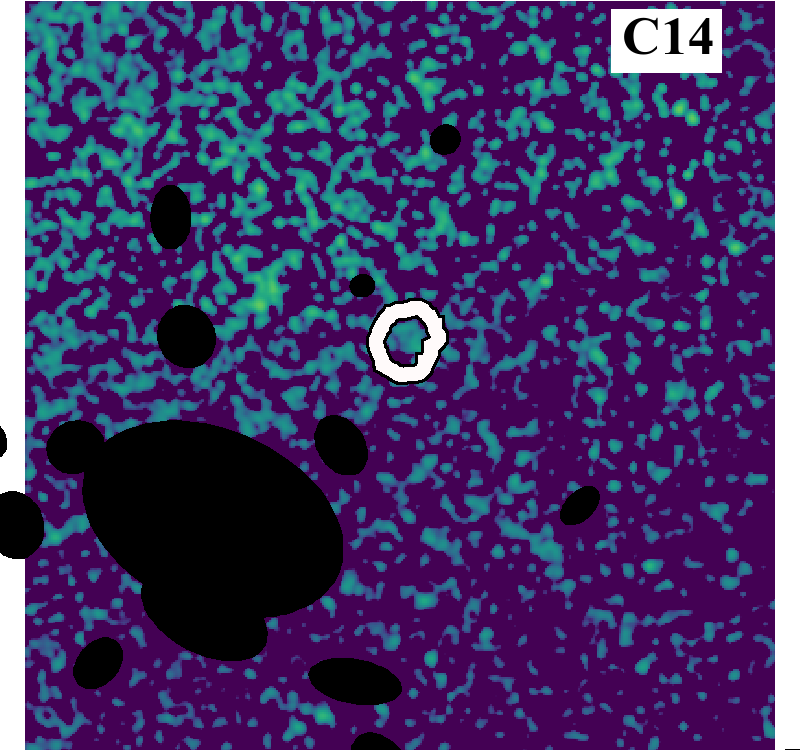}
     \includegraphics[trim=2cm 2cm 2cm 0cm, clip, width=.245\textwidth]{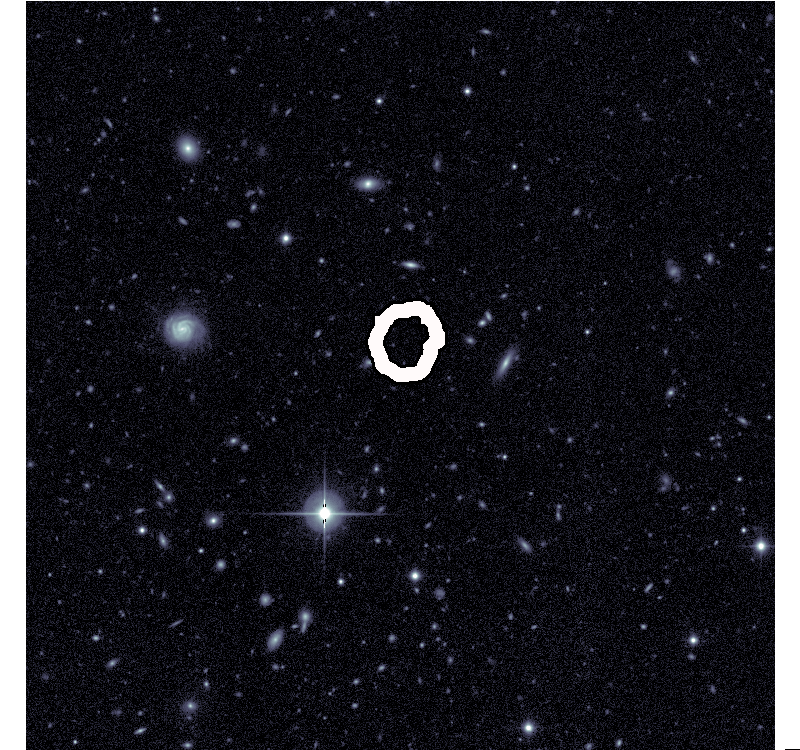}
      \hspace{.25cm} 
    \includegraphics[trim=2cm 2cm 2cm 0cm, clip, width=.245\textwidth]{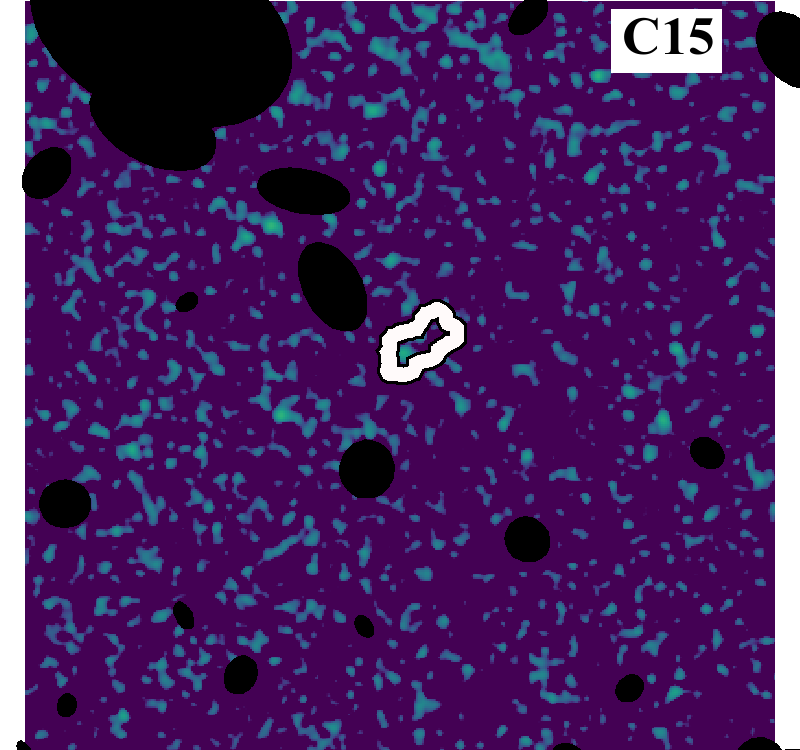}
     \includegraphics[trim=2cm 2cm 2cm 0cm, clip, width=.245\textwidth]{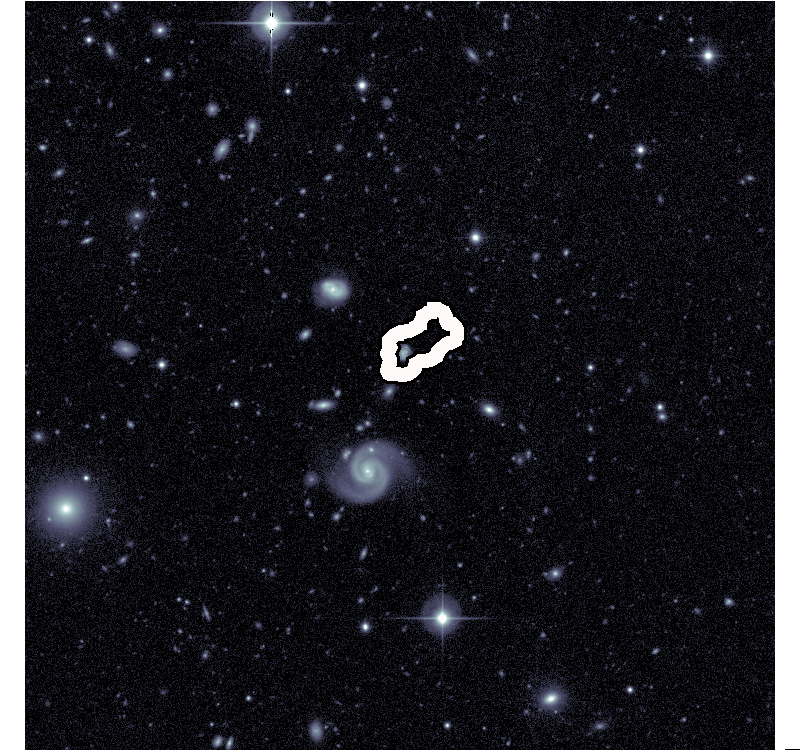}
     \includegraphics[trim=2cm 2cm 2cm 0cm, clip, width=.245\textwidth]{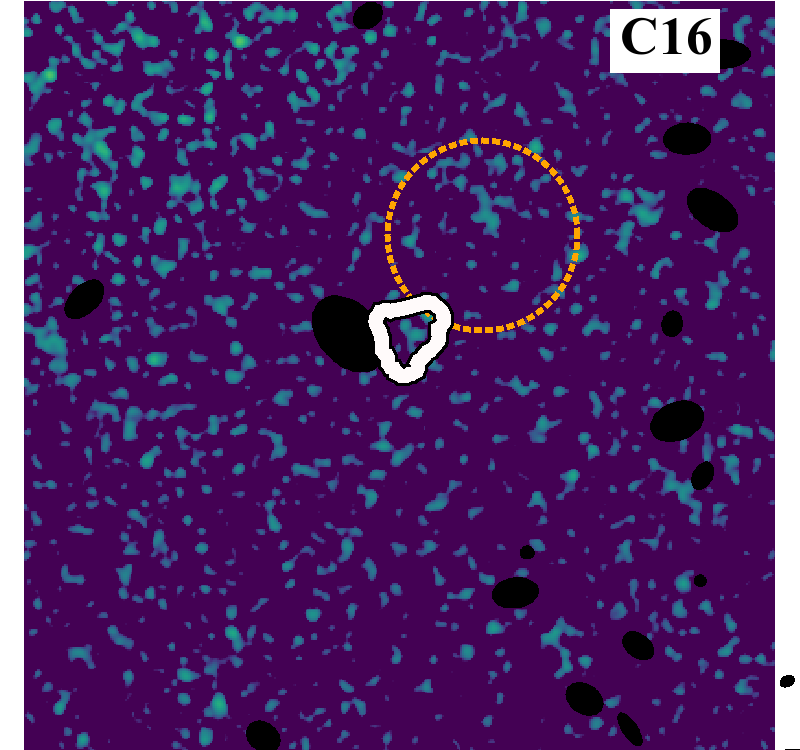}
     \includegraphics[trim=2cm 2cm 2cm 0cm, clip, width=.245\textwidth]{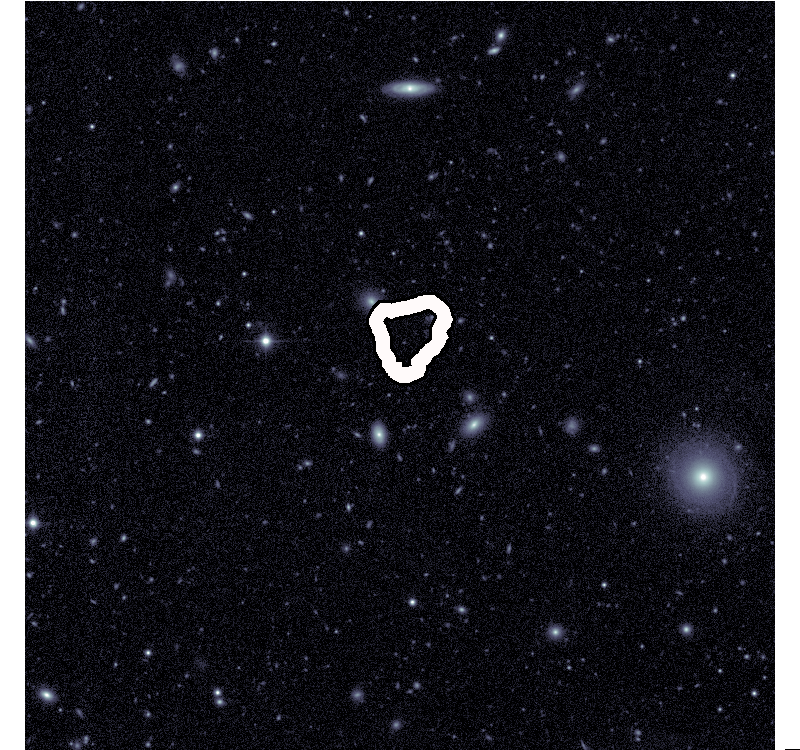}         
     \caption{Continued. }

    \label{fig:clumps}
\end{figure*}

\begin{figure}
    \centering
      \includegraphics[width=0.49\textwidth]{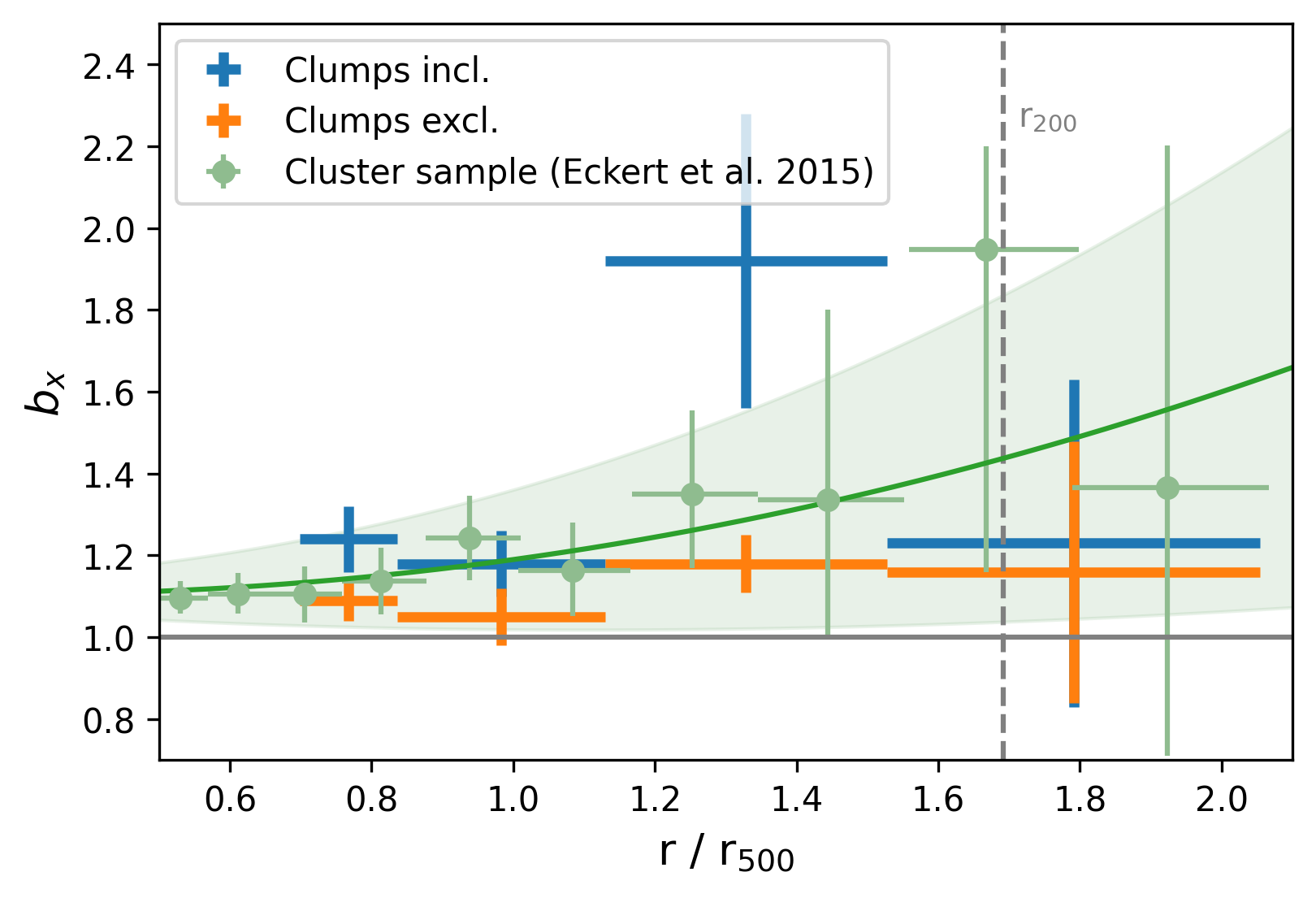}
    \caption{The azimuthally averaged emissivity bias profiles before (blue) and after (orange) removal of the identified clumps. The overplotted green points and shaded region denote the measurements obtained for a sample of 31 clusters and their best-fit profile using a second order polynomial, as reported in \citet{Eckert2015}. } 
    \label{fig:bx}
\end{figure}


\section{Spectral analysis}
\label{sec:spec}
To study the thermodynamic structure of Abell 133, we performed spectroscopic analysis with {\it Suzaku} data.
We first divided the entire region into four azimuths (NW, NE, SW and SE) and for each azimuth, we divided the observed region from 5$\arcmin$ to 30$\arcmin$ into five partial annuli, i.e., 5$\arcmin$-8$\arcmin$, 8$\arcmin$-11$\arcmin$, 11$\arcmin$-15$\arcmin$, 15$\arcmin$-20$\arcmin$,20$\arcmin$-30$\arcmin$.
The first two annuli have a width of 3$\arcmin$ each, which are practically limited by the {\it Suzaku} PSF, while the outer annuli were set wider to improve the signal-to-noise ratio (S/N).
The spectra have been extracted from individual subdivided regions.
Redistribution matrix files (RMFs) of the XIS were produced in the standard manner using {\it xisrmfgen}, and auxiliary response files (ARFs) by ray-tracing simulations using {\it xissimarfgen} \citep{Ishisaki2007}.
For all the spectral analysis, we adopted the total column density N$_{H}$ = $1.74 \times 10^{20}$~cm$^{2}$ (\citealp{Willingale2013})\footnote{Online calculator for Galactic column density: https://www.swift.ac.uk/analysis/nhtot/index.php}.
The abundance is fixed to 0.3 Solar beyond $r_{500}$ (\citealp{Werner2013Na,Urban2017}) with the abundance table from \citet{Lodders2009} applied. 

\subsection{X-ray background modeling}
\label{sec:cxb_model}
Many previous studies of the ICM in the outskirts are only based on {\it Suzaku} observations or with shallow {\it Chandra} exposure aiming to exclude bright point sources. 
However, for Abell 133 and Abell 1795, we have deep multi-telescope coverage with {\it Chandra} and {\it Suzaku}, providing a unique opportunity to test how the thermodynamic measurements in the outskirts could be improved with a combination of low-background and high imaging resolution data.
In order to investigate to what extent {\it Chandra} helps in reducing the systematic uncertainties related to the cosmic X-ray background (CXB), and how corrections for resolved clumping change the thermodynamic profiles, we have carried out three rounds of spectral analysis as introduced below.

In the first round of analysis only based on {\it Suzaku} data, hereafter R1, we only removed {\it Suzaku} detected point sources (see Figure \ref{fig:suzaku_psrc}) and estimated the CXB normalization using the outermost spectra (i.e., extracted from 20$\arcmin$-30$\arcmin$) where the ICM temperature is expected to be lower than 2.5 keV.
Since we also have to model the non--X-ray background (NXB; see Section \ref{sec:nxb_model}), this procedure needs to follow several steps. 
We first performed a fit in the 0.7-12 keV band, modelling the observations and NXB spectra in parallel, with the index and normalization of the CXB component fixed to reasonable values (e.g.,$\gamma$ = 1.52 and norm = 1.0$\times10^{-3}$~${\rm photons}~{\rm keV}^{-1} {\rm cm}^{-2}~{\rm s}^{-1}$).
During the fit, we allowed different overall norms between observations and NXB to account for the uncertainty of NXB.
After obtaining the best-fit model, we set free both CXB parameters, freeze the NXB spectral model and fit only the 4-7 keV band where the CXB emission dominates.
We found the best-fit indices for each annulus are statistically consistent with the adopted value, $\Gamma = 1.52$, therefore we fixed $\gamma$ to minimize the free parameters in the subsequent analysis. 
The corresponding best-fit CXB norm is 1.05$\times10^{-3}$~${\rm photons}~{\rm keV}^{-1} {\rm cm}^{-2}~{\rm s}^{-1}$, which we have applied for the first round of analysis.

In the second round of analysis (R2), we have further excluded the {\it Chandra} detected point sources with 2--8 keV flux above $5.67\times10^{-15}~{\rm erg}~{\rm cm}^{-2}~{\rm s}^{-1}$, which corresponds to 80\% completeness, in order to suppress the CXB variation. The remaining unresolved CXB flux has been estimated using {\it cxbtool} (\citealp{Mernier2015, Jelle2017}), by integrating the derivative source luminosity function (dN/dS) fitted based on Chandra Deep Field South (CDFS; \citealp{Lehmer2012}) data.  
The integration gives an unresolved 2--8 keV flux of (8.77$\pm$0.08) $\times 10^{-12}$ erg~cm$^{-2}$~s$^{-1}$~deg$^{-2}$.
Since we use exclusion radii r $\sim$1$\arcmin$, which corresponds to the half-power diameter (HPD) of the {\it Suzaku} PSF, 50\% of the total flux of the excluded point sources, i.e. 2.02$\times 10^{-12}$ erg~cm$^{-2}$~s$^{-1}$~deg$^{-2}$, will have been scattered into our regions of interest. We have therefore added this component to our model. 
When fitting the {\it Suzaku} spectra, we fixed the slope of the {\it powerlaw} model to $\Gamma=1.4$ and normalization to 7.57$\times10^{-4}$ ~${\rm photons}~{\rm keV}^{-1} {\rm cm}^{-2}~{\rm s}^{-1}$.

To explore how clumps affect the thermodynamic profiles, we have carried out the third round of spectral analysis (R3) by further excluding all the remaining clump candidates and utilizing the same CXB component as applied in the second round. 
We note that half of the 16 {\it Chandra}-selected clumps were not part of the {\it Suzaku} spectral analysis, either because they were outside of the field of view of the mosaic (BC1, BC2, C12 and C15), or because they are too close to point sources identified with {\it Suzaku} (C9, C11, C13 and C14).
After masking {\it Chandra} point sources in R2, two more clumps (C6 and C10) were inevitably removed.
Therefore we further removed 6 clumps (C3, C4, C5, C7, C8 and C16) in the third round of spectral analysis compared to R2.



\subsection{X-ray foreground modeling}
\label{sec:fxb_model}
The X-ray foreground spectral model includes two thermal components modeling the Galactic halo (GH, \citealp{Kuntz2000}), and the local hot bubble (LHB, \citealp{Sidher1996}) respectively.
To estimate the contribution of the GH and LHB, we obtained {\it ROSAT All-sky Survey} (RASS) data in an annulus between 1.5r$_{200}$ and 2.5r$_{200}$ around A133 with the ROSAT X-Ray Background Tool {\it sxrbg}\footnote{https://heasarc.gsfc.nasa.gov/cgi-bin/Tools/xraybg/xraybg.pl} \citep{Sabol2019}. 
In XSPEC, we fitted the background spectrum in the 0.1--2 keV band with a {\it tbabs*(apec+powerlaw)+apec} model adopting AtomDB v3.09 and the abundance table from \citet{Lodders2009}.
We fixed the power-law parameters to the values reported by Kuntz \& Snowden (2000; $\Gamma$ = 1.46, and $ Y = 8.88 \times 10^{-7}~{\rm photons}~{\rm keV}^{-1} {\rm cm}^{-2}~{\rm s}^{-1}~{\rm arcmin}^{-2} $ at 1 keV), and fitted with $\chi^{2}$ statistics.
More detailed information is listed in Table \ref{tab:model}.

Despite the fact that Abell 133 is located quite far from the Galactic Plane, where the $\sim$0.6 keV hot foreground (HF, \citealp{Yoshino2009}) originates, we checked the potential influence by adding this component. 
The $\chi^2$ statistics is not improved with this component included.
Using {\it ftest} from the heasoft package, we calculate the F-statistic and its probability given the $\chi^2$ values and the corresponding degrees of freedom (DOF) before and after adding this component.
The obtained probability of 0.36 is not low enough to indicate the necessity of adding a new component.

\begin{table*}
\begin{center}
\caption{Spectral fitting models and parameters.}
\label{tab:model}
\begin{tabular}{l | c c c }
\hline\hline
Components & \multicolumn{3}{c}{XSPEC}  \\
& Model & {\it kT} & Norm\tablefootmark{a}  \\ \hline\hline
ICM & {\it tbabs*apec} &  -- & --  \\
LHB & {\it apec} & $ 9.25\times10^{-2} $ & $6.88\times10^{-4}$    \\
GH & {\it tbabs*apec} & $0.198$   & $1.37\times10^{-3}$  \\[5pt] \hline
& Model & $\Gamma$ & Norm\tablefootmark{b}  \\ \hline\hline
CXB (R1)& {\it tbabs*pow} & 1.52 & $1.05\times10^{-3}$\\
CXB (R2\&R3)& {\it tbabs*pow} & 1.40 & $7.57\times10^{-4}$\\

\hline
\end{tabular}
\end{center}
\tablefoot{The temperatures are given in keV. The column density is $1.74\times10^{20}$~cm$^{-2}$.} 
\tablefoottext{a}{normalisation of {\it apec} model defined as $\frac{10^{-14}}{4\pi[D_{A}(1+z)]^{2}}  \int n_{\rm e} n_{\rm H} dV$ in units of cm$^{-5}$, where $D_{A}$ is the angular distance and $n_{\rm e}$, $n_{\rm H}$ represent the electron and hydrogen density. } \\
\tablefoottext{b}{the two different CXB normalisations used for the different rounds of analysis. The unit is photons~keV$^{-1}$~cm$^{-2}$~s$^{-1}$ at 1 keV. All normalizations are calculated assuming uniform emission from a circular region with a radius of 20 arcminutes.}
\vspace{10pt}
\end{table*}

\subsection{Particle background modeling}
\label{sec:nxb_model}
We followed the procedures described in \citet{Zhu2021} to generate the particle background spectra and model them later in the spectral analysis.
In brief, we created the NXB spectrum of each XIS sensor using {\it xisnxbgen}, which extracts spectra from the integrated night-Earth data collected during the period of $\pm$150 days from the target observation.
Based on Fig.1 of \citet{Tawa2008}, we modeled the NXB spectra of the FI CCDs (XIS 0 and XIS 3) with a power-law for the continuum and nine gaussian components for the instrumental lines.
For the BI CCD (XIS1), we added another broad gaussian model to account for the continuum bump above 7 keV.
More details can be found in Section 2.5 and Appendix D of \citet{Zhu2021}.

\section{Thermodynamic profiles}
\label{sec:thermo}
\subsection{ {\it Suzaku} projected profiles}
\begin{figure*}
    \centering
       \includegraphics[width=0.45\textwidth]{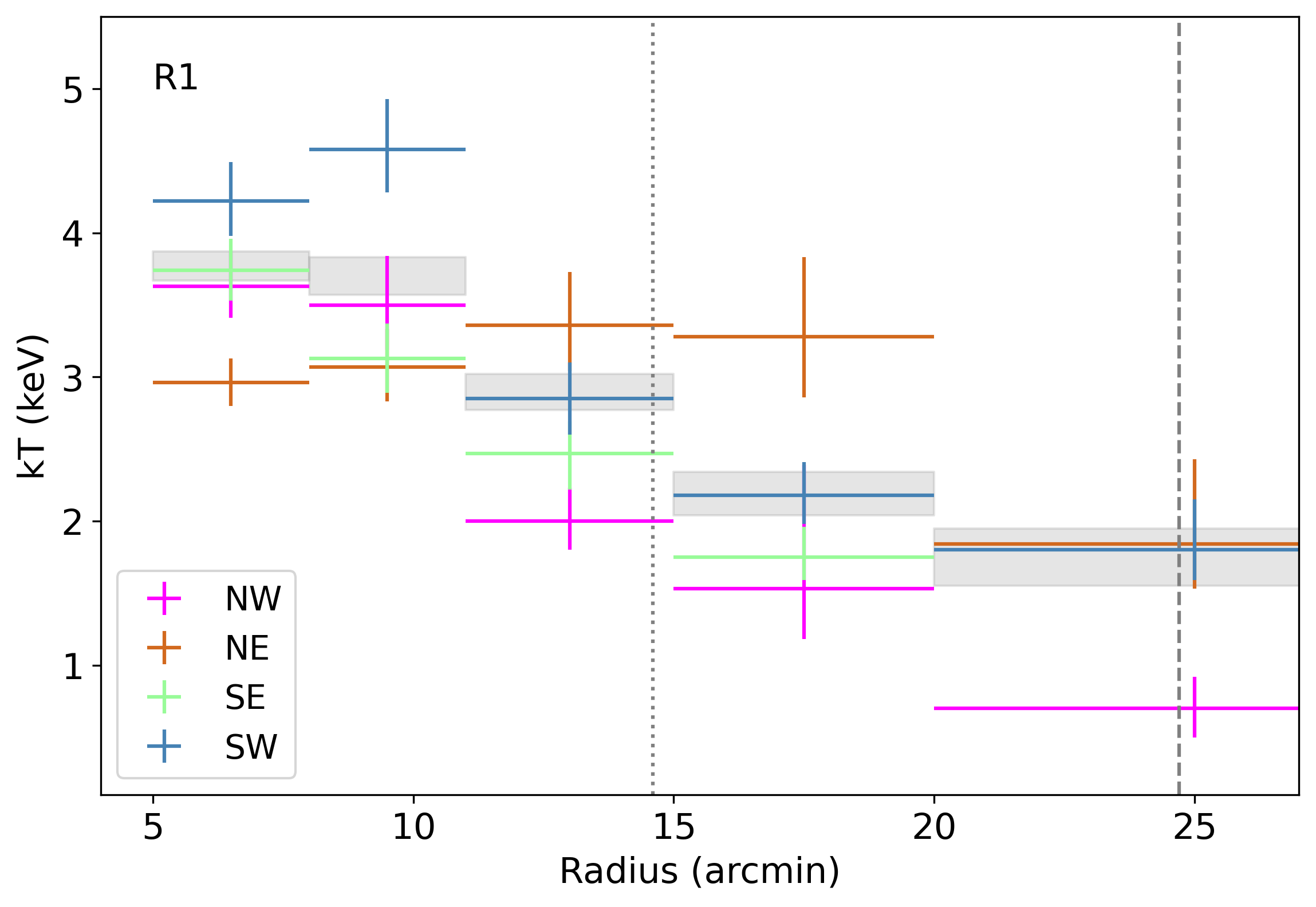}
       \includegraphics[width=0.46\textwidth]{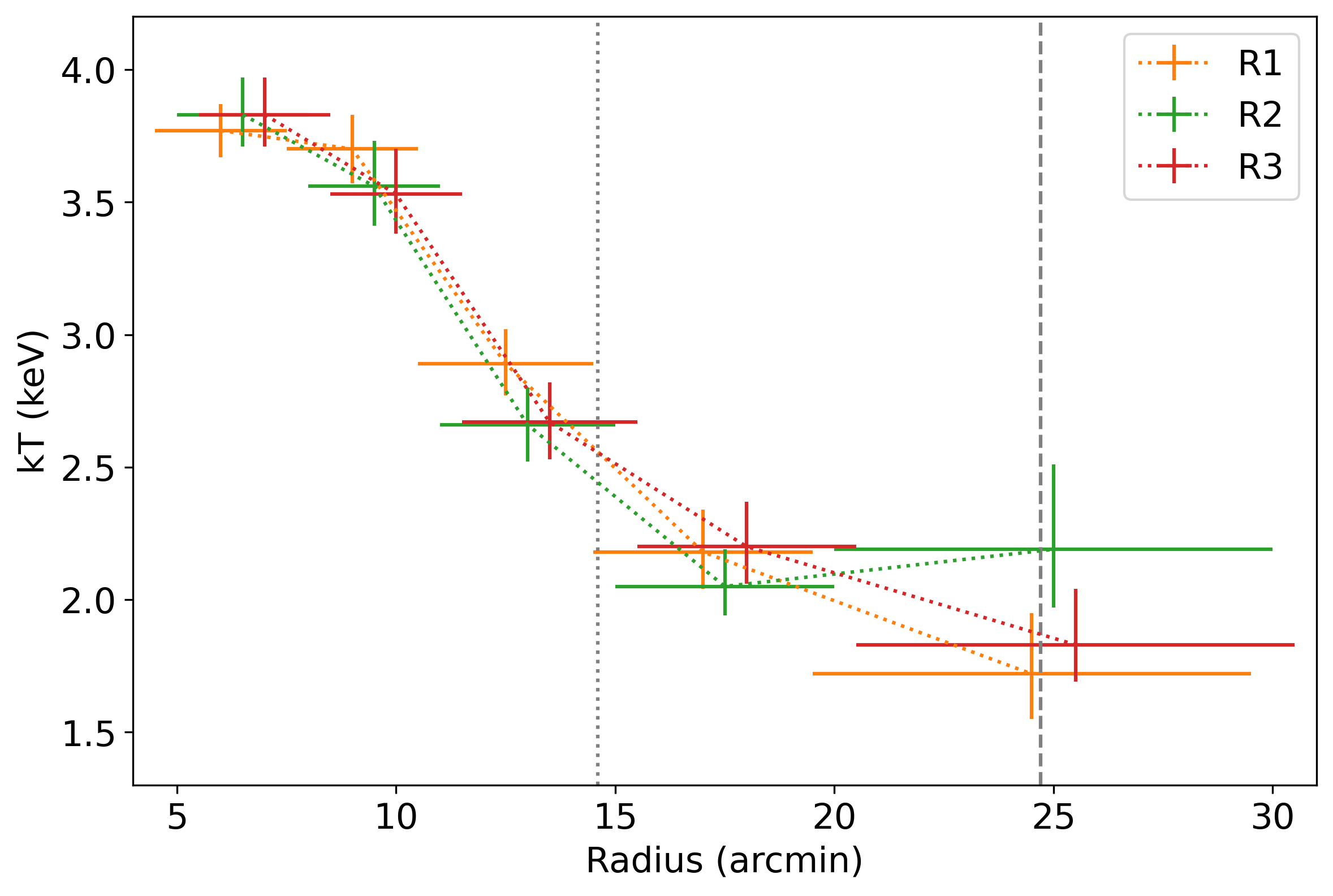}
     \caption{Projected temperature profiles. {\it Left}: The first round results (R1), using only the \textit{Suzaku} data. The grey shaded regions denote the azimuthally-averaged temperature measurements. The dotted and dashed lines show the radius of $r_{500}$ and $r_{200}$ respectively. {\it Right}: Comparison between different rounds of azimuthally averaged measurements. All three rounds of spectra were extracted from the same annuli but shifted by 0.5$\arcmin$ in this plot for illustration purpose. The abundance is fixed to 0.3 beyond $r_{500}$ (\citealp{Werner2013Na,Urban2017}). 
    }
    \label{fig:kt_proj}
\end{figure*}

The projected temperature profiles of four different azimuths obtained from the analysis described in Section \ref{sec:spec} are shown in the left panel of Fig \ref{fig:kt_proj}.
The scatter of the measurements along different azimuths indicates the inhomogeneity/asymmetry of the ICM.
There is no signal measured between 20$\arcmin$--30$\arcmin$ towards the southeast (SE).
The measured temperatures in the northwest (NW) and southeast (SE) are overall lower compared to the other two azimuths, which might be explained by the orientation of the major axis of Abell 133.
In the right panel of Figure \ref{fig:kt_proj} we present the azimuthally averaged projected temperature profiles, and compare the results obtained with the several different analysis rounds introduced in Section \ref{sec:cxb_model}.
The profiles are generally in good agreement for all rounds of analysis.

\subsection{Deprojected spectral results}
To obtain the deprojected properties, we have applied the XSPEC model {\it projct} assuming spherical symmetry.
With {\it projct}, the ICM in each annulus is modelled as the superposition of the ICM from a shell corresponding to that annulus, with the emission projected on to that annulus from the shells exterior to it.
The emission from each shell is described by an absorbed {\it apec} model.
The deprojected densities are derived from the {\it apec} normalizations for each annulus obtained during the spectral fitting, using the volumes of each shell and assuming the plasma is fully ionized with $n_{e}: n_{H} = 1.2 : 1$.
We generated a background spectrum based on the best-fit NXB model using {\it fakeit}, setting the exposure the same as the original NXB spectrum produced by {\it xisnxbgen}, to ensure the statistical uncertainties of the NXB spectrum have been properly propagated while the applied background spectrum is smooth and free from outliers in the initial background data. 
Based on the resulting measurements of deprojected temperatures and electron density, we further derived the deprojected pressure ($P = n_{e} kT$) and entropy ($K = kT/n_{e}^{2/3}$).
In Figure \ref{fig:KP}, we compare the azimuthally averaged entropy and pressure profiles with their reference models described below.

\subsubsection{Entropy Profile}
In a cluster formed by gravitational collapse without additional heating or cooling, the entropy is expected to follow a power-law: \\
\begin{equation}
 K/K_{500} = 1.47 (r/r_{500})^{1.1} , 
\end{equation}
where $K_{500} = 106$~keV~cm$^{2} (M_{500} / 10^{14} M_{\odot})^{2/3} E(z)^{-2/3}$ (\citealp{Voit2005, Pratt2010})\footnote{$E(z) = \sqrt{\Omega_{m}(1+z)^{3}+\Omega_{\Lambda}}$ is the ratio of the Hubble constant at redshift $z$ with its present value.}. 
We adopted $f_{b}$ = 0.15, $r_{500}$ = 14.6$\arcmin$ and $M_{500} = 3.2\times10^{14} M_{\odot}$ \citep{Vikhlinin2006} in calculating $K_{500}$.

As shown in Figure \ref{fig:KP}, beyond $\sim$0.7$r_{200}$ we observed a flattening of the entropy profile with respect to this expected power-law for all three rounds of analysis.
We overplotted the model introduced by \citet{Walker2012b}, who proposed that the entropy profile outside of 0.2$r_{200}$ can be fitted with an analytical function:\\
\begin{equation}
K/K (0.3r_{200}) = A (r/r_{200})^{1.1} {\rm exp}[ - (r/Br_{200})^{2} ],
\end{equation}
with (A, B) = $(4.4^{+0.3}_{-0.1}, 1.00^{+0.03}_{-0.06})$.
The entropy profiles of all three rounds, in particular the outermost measurement, indeed achieved a better match with this second model.

\subsubsection{Pressure Profile}
We utilized the generalized NFW pressure profile proposed by \citet{Nagai2007} as a reference model, which follows the form:\\
\begin{equation}
\centering
\frac{P(r) }{ P_{500} } = \frac{ P_{0}  }{(c_{500}x)^{\gamma} [1+(c_{500}x)^{\alpha}]^{(\beta-\gamma)/\alpha}  },
\end{equation}
where $x = r/R_{500}$, $P_{0}$ is the normalization, $c_{500}$ is the concentration parameter defined at $r_{500}$, and the indices $\alpha$, $\beta$ and $\gamma$ are the profile slopes in the intermediate, outer and central region.
The characteristic pressure $P_{500}$ scales with the cluster mass:
\begin{equation}
P_{500} = 1.65\times10^{-3} E(z)^{8/3}\times \left[ \frac{M_{500}}{3\times10^{14} h_{70}^{-1}M_{\odot}} \right] ^{2/3}~ h_{70}^{2} {\rm keV}~{\rm cm}^{-3}.
\end{equation}
\citet{Arnaud2010} gave the best-fit parameters, $(P_{0}, c_{500}, \alpha, \beta, \gamma)_{A}$= (8.403, 1.177, 1.0510, 5.4905, 0.3081), by fitting 33 nearby (z < 0.2) clusters out to $0.6r_{200}$. 
By analyzing 62 Planck clusters between 0.02$r_{500}$ < r < 3$r_{500}$, \citet{Planck2013} obtained another set of parameters as $(P_{0}, c_{500}, \alpha, \beta, \gamma)_{P}$ = (6.41, 1.81, 1.33, 4.13, 0.31).
As shown in the lower panel of Figure \ref{fig:KP}, our measurements show a reasonable agreement with the Planck measurements, though the outermost data point appears marginally higher.

\begin{figure*}
    \centering
     \includegraphics[width=0.94\textwidth]{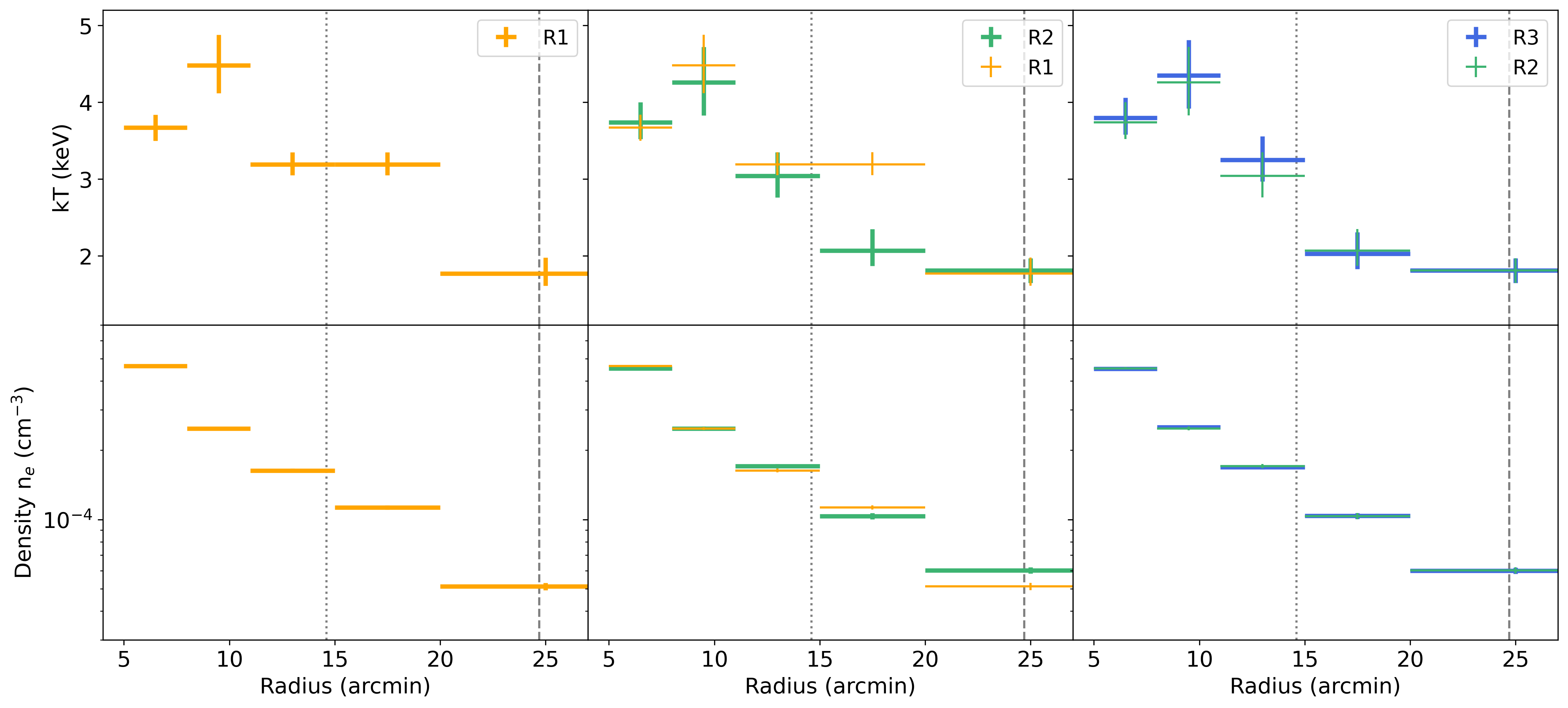}
    \caption{Deprojected temperature and density profiles. {\it Left:} {\it Suzaku} only measurements (R1) denoted with orange data points. {\it Middle:} Measurements using the {\it Chandra} point source list (R2) shown as green data points, with R1 results overplotted. {\it Right:} The comparison between profiles before (R2) and after (R3) the removal of {\it Chandra}-detected clumps. 
    }
    \label{fig:Tne}
\end{figure*}

\begin{figure*}
    \centering
    \includegraphics[width=0.94\textwidth]{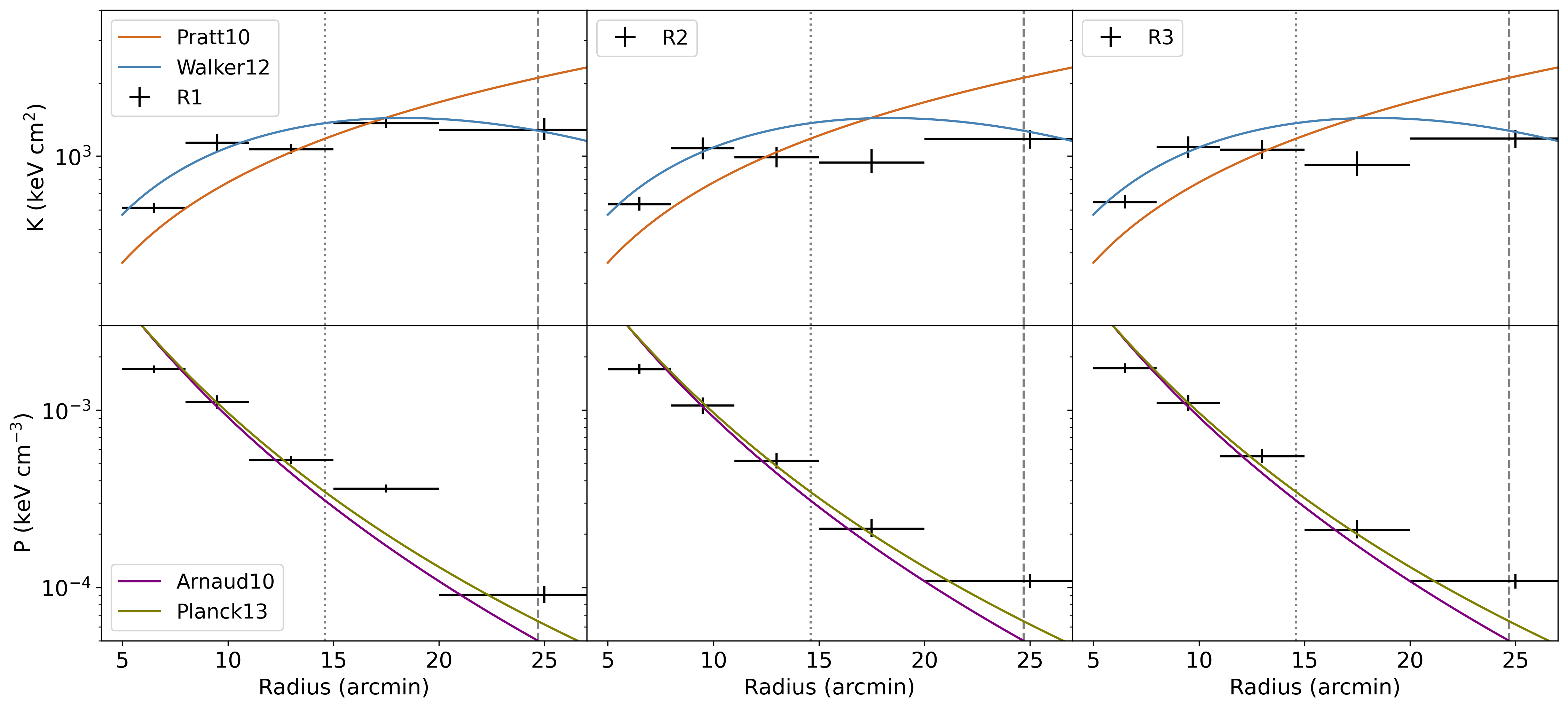}
    \caption{Azimuthally averaged profiles of the ICM properties. {\it Top:} Entropy profile. The orange line shows the entropy model calculated according to \citet{Pratt2010}. The blue line indicates the best-fit entropy profile by \citet{Walker2012b}. {\it Bottom:} Pressure profile with overplotted models by \citet{Arnaud2010} (purple line) and \citet{Planck2013} (olive line). The dotted and dashed vertical lines show the radius of $r_{500}$ and $r_{200}$, respectively. 
    }
    \label{fig:KP}
\end{figure*}

\section{Discussion}
\label{sec:discussion}
To understand how the addition of complementary high-spatial resolution {\it Chandra} data improves our understanding of the ICM measurements, we have carefully compared the results between Suzaku-based (R1) and Chandra-involved (R2\&R3) analysis.
As shown in the right panel of Fig \ref{fig:kt_proj}, the projected temperatures measured in different rounds are roughly consistent within statistical uncertainties. 
The deprojected profiles show a  difference in the 15$\arcmin$-20$\arcmin$ annulus between R1 and R2 results. However, this is simply due to the fact that for the {\it Suzaku} only analysis, we were forced to couple the ICM temperature of the 15$\arcmin$-20$\arcmin$ annulus to that of the 10$\arcmin$-15$\arcmin$ annulus when fitting with {\it projct}. 
With {\it Chandra}-selected point sources removed, R2 and R3 have more precise constraints on the CXB level and the cosmic variance has been suppressed; thus the deprojection was more stable and this coupling of temperatures between neighboring annuli was no longer needed.

\subsection{Systematic Uncertainties}
Spatial variations in the foreground emission and cosmic variance can introduce systematic uncertainties in the measured thermodynamic profiles.
To estimate the variation of GH, we have extracted RASS spectra of circles with radius of 1 degree, in four different azimuths outside 1.5$r_{200}$, and calculated the variance in the best-fit normalization of GH.    
The obtained 27\% systematic uncertainty for GH is then taken into account.
The expected cosmic variance due to unresolved point sources over a solid angle $\Omega$ is 
\begin{equation}
\sigma_{B}^{2} = \frac{1}{\Omega} \int_{0}^{S_{\rm excl}} \frac{dN}{dS} \times  S^{2} dS,
\end{equation}
where $\Omega$ is the solid angle (\citealp{Bautz2009}). We substituted the derivative source function (dN/dS) given in \citet{Lehmer2012} and adopted a flux cut of $S_{\rm excl}$ = 5.67$\times 10^{-15}$ erg~cm$^{-2}$~s$^{-1}$ (See Section \ref{sec:cxb_model}) for R3.
We found the relative cosmic variance $\sigma_{B}/B$ in the outermost annulus decreases from 6\% in R1 to 3\% in R3.

\begin{figure}
    \centering
        \includegraphics[width=0.48\textwidth]{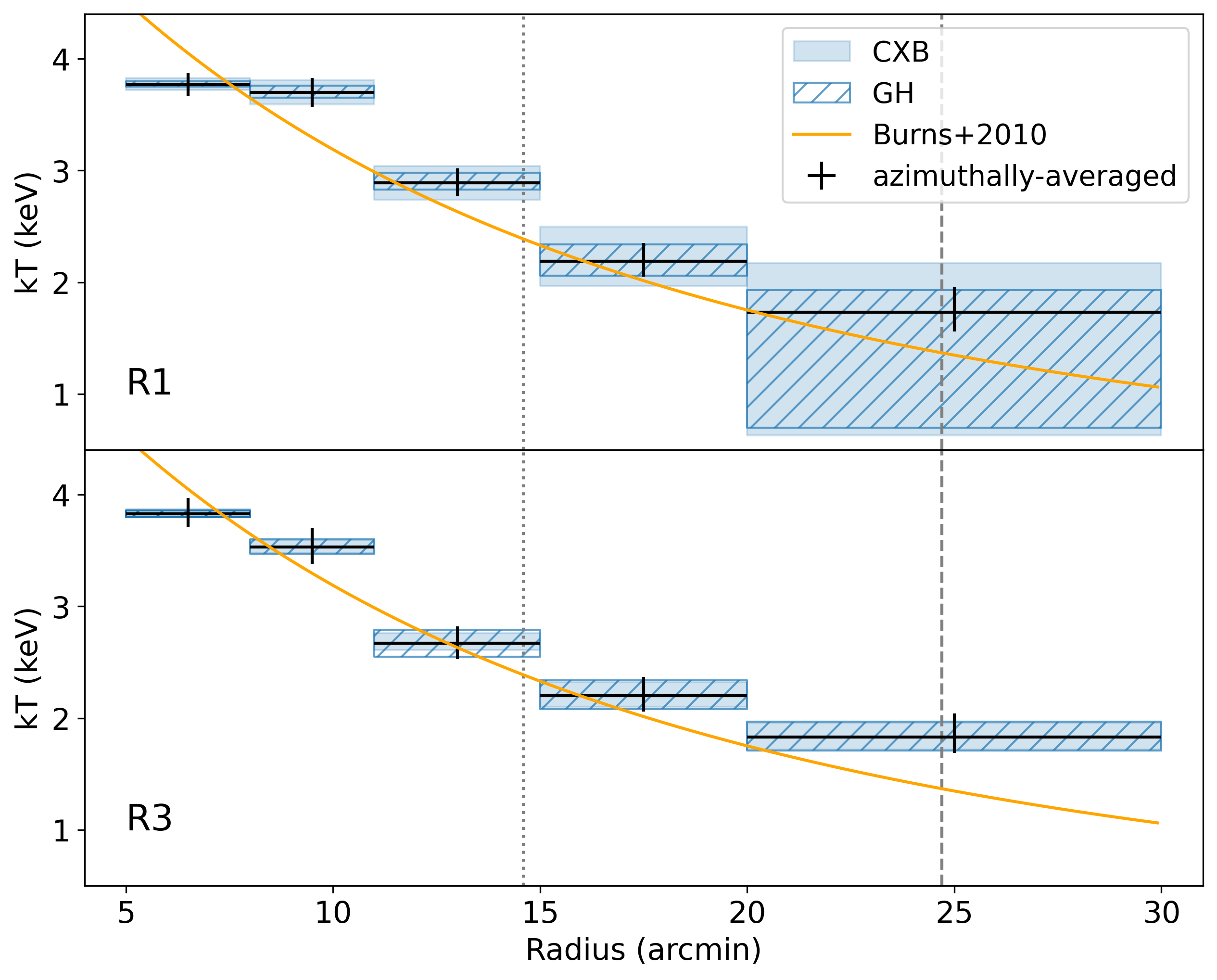}
    \caption{{\it Suzaku} temperature profile of Abell 133. Systematic uncertainties from best-fit GH model and CXB estimation are denoted by blue slash region and shaded region respectively. The upper and lower panels show the differences before and after removal of point sources (S$_{2-8}$ > $5.67\times10^{-15}~{\rm erg}~{\rm cm}^{-2}~{\rm s}^{-1}$  ) and clumps. The orange solid line represents the expectation from hydrodynamic cosmological simulations \citep{Burns2010}. } 
    \label{fig:kt_avg}
\end{figure}

In Figure \ref{fig:kt_avg}, we demonstrate the impact of systematic uncertainties on the azimuthally-averaged temperatures for R1 and R3.
For the {\it Suzaku}-only analysis shown in the upper panel, it is noteworthy that while systematic errors are comparable with the statistical errors for measurements within r$_{500}$, they become increasingly larger in the outskirts especially approaching r$_{200}$.
It also shows that outside 1.4r$_{500}$ thermodynamic measurement results become extremely sensitive to the cosmic variance, while for the inner regions, they are much more robust.
For the outermost bin, systematic uncertainties from the GH emission have also increased likely due to model degeneracies causing the best fit ICM component to compensate for a combination of GH and CXB residuals.
The measurements roughly match with the expectation from hydrodynamic cosmological simulations \citep{Burns2010}, taking into account the large systematic error on the outermost data point.
Through comparing the upper (R1) and lower (R3) panels of Figure \ref{fig:kt_avg}, we show that the systematic uncertainties have been significantly reduced after the removal of more point sources benefiting from the {\it Chandra} observations.

\subsection{Clumping correction}
In the lower panels of Figure \ref{fig:Tne}, we compare the density profiles of different rounds.
We have applied power-law modelling for the density profiles of the outskirts of Abell 133, $n_{e} = n_{0}~r^{-\delta}$.

We report a relatively flat azimuthally averaged density profile of R2, falling off with radius with an index of $\delta= 1.60 \pm 0.06$ outside 0.6 Mpc (r > 0.4~$r_{200}$), in agreement with the slope reported by \citet{Urban2014} for the Perseus Cluster in the same scaled radial range.
After removing the resolved clumps, the best-fit density slope of R3 outside 0.4~$r_{200}$ remains the same as R2. 
However, as shown in Figure \ref{fig:KP}, even after the correction of resolved clumps, the entropy profile approaching the outskirts still flattens, significantly deviating from the power law model (\citealp{Pratt2010}).

\begin{figure*}
  \includegraphics[width=0.96\textwidth]{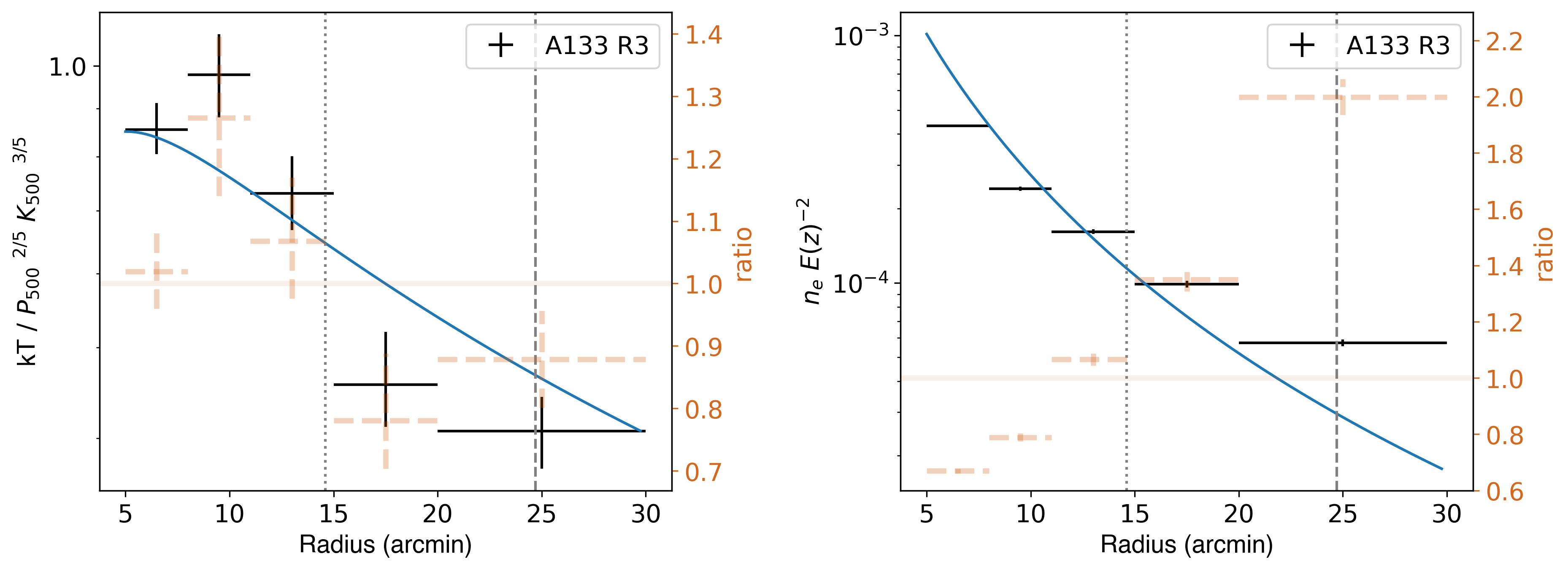}
    \caption{The self-similarly scaled temperature and density profiles of Abell 133. The dotted and dashed lines show the radius of r$_{500}$ and r$_{200}$, respectively. P$_{500}$ and K$_{500}$ are defined in Section \ref{sec:thermo}.The reference model derived from the referenced pressure and entropy profiles is denoted with a blue solid curve. The orange dashed points represent the ratio between the measurements and the reference models, for better comparison.}
    \label{fig:kt_ne_ref}
\end{figure*}

By integrating equation (1) with equation (3) (utilizing the Planck Collaboration 2013 parameters for equation 3), we can solve for the expected $n_{e}$ and $kT$ profiles. 
In Figure \ref{fig:kt_ne_ref} we compare the measured density and temperature with these baseline expectations.
We found the $n_{e}$ measured from R3, which has been best corrected for the clumping, still deviate from the baseline outside $r_{500}$. 
The ratio between the measured and predicted $n_{e}$ progressively increases towards the outskirts, ultimately reaching a value of 2.0 in the outermost bin. 

\subsection{Non-thermal pressure or electron-ion non-equilibrium}

In addition to the remaining, unresolved clumping bias, there is also a temperature bias in the outskirts, namely the measurements appear lower than the expected values.
Here, we discuss two explanations.

Besides clumping, other ways to explain the entropy deficit seen in the cluster outskirts involve scenarios wherein a sizeable fraction of the gravitational energy due to infall is not thermalised immediately.
For example, \citet{Ghirardini2018} found that the entropy profile of Abell 2319 measured from the joint analysis of the X-ray and Sunyaev-Zel'dovich (SZ) signals still shows a deficit near r$_{200}$, even after correcting for gas clumping. They attribute this to a level of non-thermal pressure due to gas motions and turbulence, amounting to 40\% of the thermal pressure near the virial radius. Similar physics could be responsible for the entropy (and temperature) deficit in the outskirts of A133, especially given its prominent radio relics indicating recent merging activity. Quantitative estimation of the proportion of non-thermal pressure however requires hydrostatic mass modelling, which is beyond the scope of this work.


Alternatively, the relatively low kT measured in the outskirts could result from non-equilibrium.
When a plasma of electrons and ions travels through a shock, most of the kinetic energy goes into heating the heavier ions, causing T$_{i}$ >> T$_{e}$.
However, with CCD spectroscopy, we can only measure the electron temperature, which may represent an underestimation of the overall gas temperature.
After the shock, electrons and ions slowly equilibrate over a typical timescale, which is 
given by 
\begin{equation}
    t_{ei} \approx 6.3\times 10^{8} yr \frac{ (T_{e}/10^{7} K)^{3/2} }{ (n_{i}/10^{-5} cm^{-3})(ln \Lambda / 40) }
\end{equation}
\citep{Spitzer1962}.
We substitute the T$_{e}$ and n$_{i}$ of the outermost annulus from R3 results into this equation and obtain a timescale of 3.8$\times$10$^{8}$ years. 
The past accretion event could be more recent compared to this timescale, causing the measured temperature to be lower than expectations.

\section{Conclusion}
\label{sec:conclusion}
We have explored {\it Suzaku} and {\it Chandra} data of Abell 133, one of the galaxy clusters with the best X-ray coverage of its outskirts region.
After removing point sources, we identified 16 clump candidates with at least 2$\sigma$ significance from the {\it Chandra} image.
Combining the cluster catalogue from the DESI Legacy Imaging Surveys and the r-band image taken with CFHT, we have discussed the origin of individual clump candidates and found that {\it Chandra}-selected clumps are mainly background clusters or galaxies, instead of genuine inhomogeneity.
In this work, we performed three rounds of {\it Suzaku} spectral analysis and derived the thermodynamic profiles to large radii of the Abell 133.
We have further compared thermodynamic profiles after the correction for resolved clumps by removing them from spectral extraction.
In general, none of the thermodynamic profiles is heavily affected by the corrections applied.
For the case of Abell 133, even after the correction for the clumping resolved by very deep {\it Chandra} data, we still see an entropy deficit and a density excess compared to the expectations. This suggests that the effect of unresolved clumping is potentially important and must still be taken into account, even when analysing very sensitive, high spatial resolution data.
Aside from the density bias, we also report a mild underestimation of the temperature at large radii. It is therefore possible that other physical mechanisms (e.g. non-thermal pressure/turbulence and non-equilibrium electrons in the ICM) in the outskirts may play an additional role.

\begin{acknowledgement}
We would like to thank Henk Hoekstra for access to CFHT data of Abell 133 and Konstantinos Migkas for useful discussion about {\it Suzaku} cosmic X-ray background estimation.

ZZ, AS are supported by the Netherlands Organisation for Scientific Research (NWO).
The Space Research Organization of the Netherlands (SRON) is supported financially by NWO.
OEK and NW are supported by the GA\v{C}R EXPRO grant No. 21-13491X ``Exploring the Hot Universe and Understanding Cosmic Feedback".
This research is mainly based on observations obtained from the \emph{Suzaku} satellite, a collaborative mission between the space agencies of Japan (JAXA) and the USA (NASA). 
This research has also made use of data obtained from the Chandra Data Archive and software provided by the Chandra X-ray Center (CXC) in the application package CIAO.

\end{acknowledgement}

\bibliography{ref}
\bibliographystyle{aa}

\begin{appendix}
\section{Solar proton flux variation during {\it Suzaku} observations}
\begin{figure}
    \centering
     \includegraphics[width=0.5\textwidth]{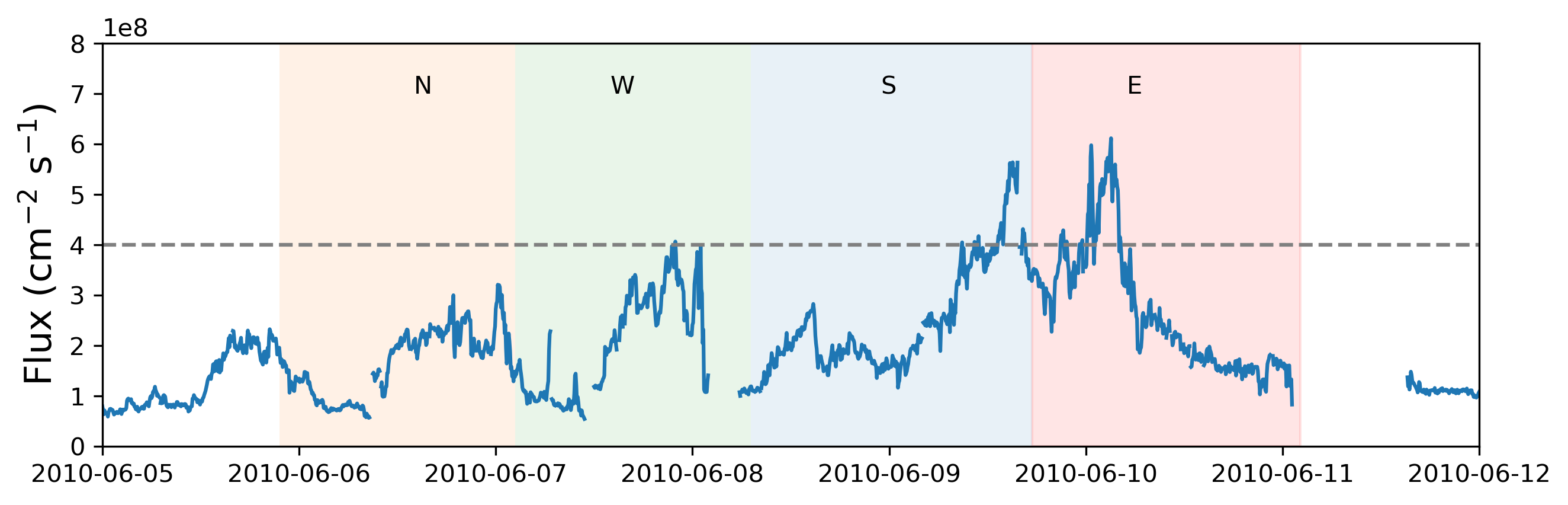}
    \includegraphics[width=0.5\textwidth]{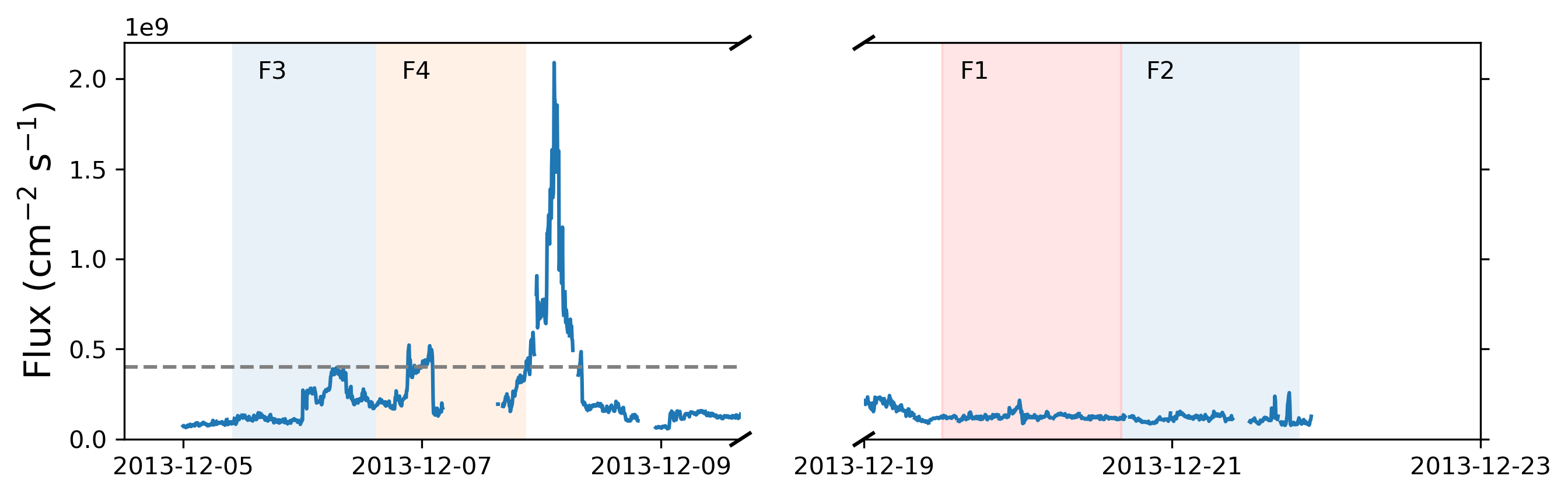}
    \caption{The solar proton flux measured by the WIND spacecraft's solar wind experiment instrument. The shaded regions denote the time coverage of {\it Suzaku} observations, corrected for the particle travel time to the earth.}
    \label{fig:swce}
\end{figure}

We checked for potential contamination from solar wind charge exchange (SWCX).
Here, in Fig.\ref{fig:swce} we plotted the solar proton flux measured by WIND spacecraft's solar wind experiment instrument during 2010 Jun 5-12 and 2013 Dec 5-9/19-23, which cover the periods of all 8 {\it Suzaku} observations used in our analysis.
Fig.\ref{fig:swce} shows that the proton flux is below 4$\times10^{8}$ cm$^{-2}$ s$^{-1}$ for observation N, W, F1, F2, F3 and F4, therefore will not produce significant contamination \citet{Yoshino2009}.
For observation S and E, we have checked the count rate of 0.7-1.2 keV where SWCX contributes the most, and this value remains unchanged within the uncertainty range.
Therefore despite the proton flux increases, these two observations appear to be free from SWCX contamination.

\section{{\it Suzaku} selected point sources}
\begin{figure}
    \centering
     \includegraphics[width=0.42\textwidth]{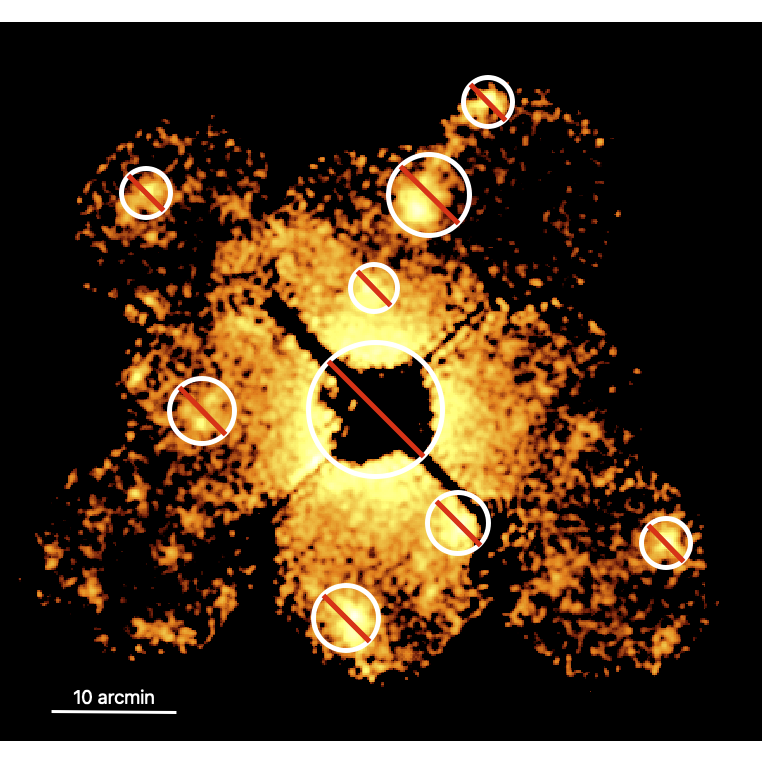}
    \caption{Exposure- and vignetting-corrected 0.7--7 keV {\it Suzaku} image of Abell 133 with the point sources that have been excluded in the {\it Suzaku}-only analysis (R1) overplotted.}
    \label{fig:suzaku_psrc}
\end{figure}

\end{appendix}

\end{document}